\documentclass[twocolumn,preprintnumbers,aps,pra,amsmath,amssymb,showpacs 
]{revtex4-1} 

\usepackage[dvips]{graphicx}
\usepackage{dcolumn}
\usepackage{multirow} 
\usepackage{amsmath,amssymb} 
\usepackage{bm}
\usepackage{color} 
\usepackage[title,titletoc,toc]{appendix} 
\usepackage{longtable} 
\usepackage[normalem]{ulem}

\begin{document} 
 
\title{An optimized absorbing potential for ultrafast, strong-field problems} 
\author{Youliang Yu} 
\author{B. D. Esry} 
\affiliation{J.R. Macdonald Laboratory, Kansas State University, Manhattan, Kansas, 
66506}  

\begin{abstract} 

Theoretical treatments of strong-field physics have long relied on the numerical 
solution of the time-dependent Schr\"odinger equation. The most effective such treatments utilize a discrete spatial 
representation\,---\,a grid. Since most strong-field observables relate to the continuum portion of the wave function, the 
boundaries of the grid\,---\,which act as hard walls and thus cause reflection\,---\,can substantially impact the observables. 
Special care thus needs to be taken. While there exist a number of attempts to solve this problem\,---\,e.g., complex 
absorbing potentials and masking functions, exterior complex scaling, and coordinate scaling\,---\,none of them are 
completely satisfactory. The first of these is arguably the most popular, but it consumes a substantial fraction of 
the computing resources in any given calculation. Worse, this fraction grows with the dimensionality of the problem. 
And, no systematic way to design such a potential has been used in the strong-field community. In this work, we 
address these issues and find a much better solution. By comparing with previous widely used absorbing potentials, 
we find a factor of 3--4 reduction in the absorption range, given the same level of absorption over a specified 
energy interval. 

\end{abstract} 

\maketitle

\section{Introduction} 

To theoretically describe highly nonperturbative interactions\,---\,such as strong-field physics\,---\,in 
a fully quantitative manner, the best 
option is usually to numerically solve the time-dependent Schr\"odinger equation (TDSE). 
One of the most popular approaches to practically solving the TDSE represents the wave function on a finite spatial grid 
with boundary conditions applied at its edges. 
In general, such a grid needs to be large enough so that there are no 
reflections from the boundaries which behave as infinitely hard walls. Otherwise, the reflections 
might lead to unphysical changes in the observables.
For example, an ionized wavepacket reflected from  the boundary back to the origin might be driven 
by the laser field into bound states, thus reducing the total ionization yield. 

The most direct way to avoid such spurious reflections is to move the boundary further away. Since the grid 
density is fixed physically by the highest energy, however, this requires more grid points which, in 
turn, incurs a greater computational cost.  In fact, the large grids required to describe
current experiments have become 
a key bottleneck to improving the numerical efficiency of solving the TDSE, especially as
laser wavelengths push  beyond 800\,nm. 

Fortunately, if the wave function at large distances  
can  easily be  reconstructed or
is not of interest,  
it can   be absorbed at a sufficiently large distance 
that it does not affect the dynamics at small distances. 
Applying such absorption techniques, one can generally reduce the box size significantly. 
The absorb-and-reconstruct strategy was probably first developed by Heather and Metiu~\cite{Heather_Metiu_JCP_1987} which they demonstrated for strong-field
dissociation.  Their work has been adopted in hundreds of papers since.  A new implementation
following this philosophy~\cite{TSURF_NJP_2012,iSURF_Patchkovskii_JPB_2016} has proven similarly effective.

Among the various methods to effect absorption at the boundary, the most  widely used\,---\,and probably the simplest\,---\,method 
is the complex absorbing potential 
(CAP)~\cite{KOSLOFF_JCompP_1986,TISE_Balint-Kurti_JCP_1992,TDSE_Balint-Kurti_JPC_1992, Muga_CPL_1994, Riss_Meyer_JCP_1996,
Zhang_JCP_1998, Hussain_PRA_2000,Manolopoulos_JCP_2002, Poirier_JCP_2003, Muga_PhysRep_2004,
Greene_PRA_2012, Cos2CAP_PRA_2012, LunYue_PRA_2013, Bernhard_JCP_2014} 
or the closely related masking function~\cite{Kulander_MaskingFunc_PRA_1992}.  
Another increasingly popular absorbing-boundary technique is exterior complex 
scaling~\cite{McCurdy_Science_1999,TISE_McCurdy_PRA_2000,ECS_Becker_PRA_2007,ECS_McCurdy_PRA_2009, irECS_Scrinzi_PRA_2010}, 
where one rotates the coordinate into the complex plane at large distances.  
Other, less common, methods to treat the boundary reflection include time-dependent
coordinate scaling~\cite{Sidky_PRL_2000, Piraux_PRA_2011, Madsen_PRA_2016},
the interaction representation~\cite{IntPic_Zhang_JCP_1990,InteractionPic_Tannor_JCP_1991,IntRep_CPL_1992}, 
and Siegert-state expansions~\cite{Burgdorfer_PRA_1999}.
While these methods are local in time and vary from exact to approximate, it is
also possible to construct a perfectly transparent boundary using 
Feshbach projection techniques~\cite{Muga_PhysRep_2004}.
The disadvantage of such methods is that they require the wave function from previous times and are thus
nonlocal in time. 
In this work, we will focus on the CAP due to its popularity and the simplicity of its implementation.  Our goals
are to make it both more efficient and more effective.

Although the CAPs utilized in previous studies are predominantly 
polynomials~\cite{Riss_Meyer_JCP_1996,Zhang_JCP_1998, Poirier_JCP_2003, 
Muga_PhysRep_2004,Greene_PRA_2012, LunYue_PRA_2013,Bernhard_JCP_2014}, 
other types of absorbing potentials such as $\cos^2$~\cite{Cos2CAP_PRA_2012}, 
P\"oschl-Teller ($1/\cosh^2$)~\cite{KOSLOFF_JCompP_1986}, and a
pseudo-exponential [$\exp(-x^{-n})$]~\cite{TISE_Balint-Kurti_JCP_1992,TDSE_Balint-Kurti_JPC_1992} have also been  used. 
In most of these papers, 
the CAP's performance is examined by studying the dependence of the reflection $R$ and transmission $T$ coefficients on 
the energy. Riss and Meyer~\cite{Riss_Meyer_JCP_1996}, for instance, carefully investigated the properties of $R$ and $T$ for 
polynomial CAPs, finding some difficulty in treating low energies.
They characterized their optimized potential parameters in terms of
the absorbed energy ratio $E_{\rm max}/E_{\rm min}$, where $E_{\rm min}$ and $E_{\rm max}$ 
indicate the minimum and maximum energies, respectively, for which absorption exceeds a given 
value. The maximum ratio they considered,  30, is too small, however, 
for typical strong-field electronic dynamics. We will, for instance, consider $E_{\rm max}/E_{\rm min}$=500.
Vibok and Balint-Kurti~\cite{TISE_Balint-Kurti_JCP_1992,TDSE_Balint-Kurti_JPC_1992} presented a more optimal 
CAP\,---\,the $\exp(-x^{-n})$ type\,---\,for heavy particles, 
but the range of absorbed energies was still insufficient for strong-field problems.

Even though $R$ and $T$ provide a clear, quantitative measure of performance, studies of CAPs 
in strong-field problems utilizing them can hardly be found. 
Their absence is likely due to the inherent time-dependent nature  of the strong-field problem and the authors' consequent focus on wavepacket behavior, 
losing track of the connection with $R$ and $T$. In contrast, we will adopt 
$R$ as the figure of merit for designing our absorbing potentials for the strong-field problem, incorporating
it as a critical piece in our systematic CAP construction method.

The major advantage of the CAP is its simplicity.  The major disadvantage is that it has required a relatively large spatial range to  be effective, 
thus consuming non-negligible computational resources. 
In this paper, we improve the performance of the CAP and systematically design a more optimal\,---\,yet general\,---\,CAP 
for strong-field processes. 
Specifically, we provide an optimized CAP with a factor of 3--4 reduction in the absorption range compared to some widely-used 
CAPs~\cite{Muga_PhysRep_2004}. 
Our optimized CAP absorbs at a prescribed level over a large enough energy range to be useful for strong-field processes.  

To be clear, while we optimize our CAP for the strong-field problem, it can be readily adapted 
and re-optimized for other problems following the procedures we outline below.
 
\section{Theoretical background} 
\label{Method} 

\begin{figure} 
\includegraphics[width=0.95\columnwidth]{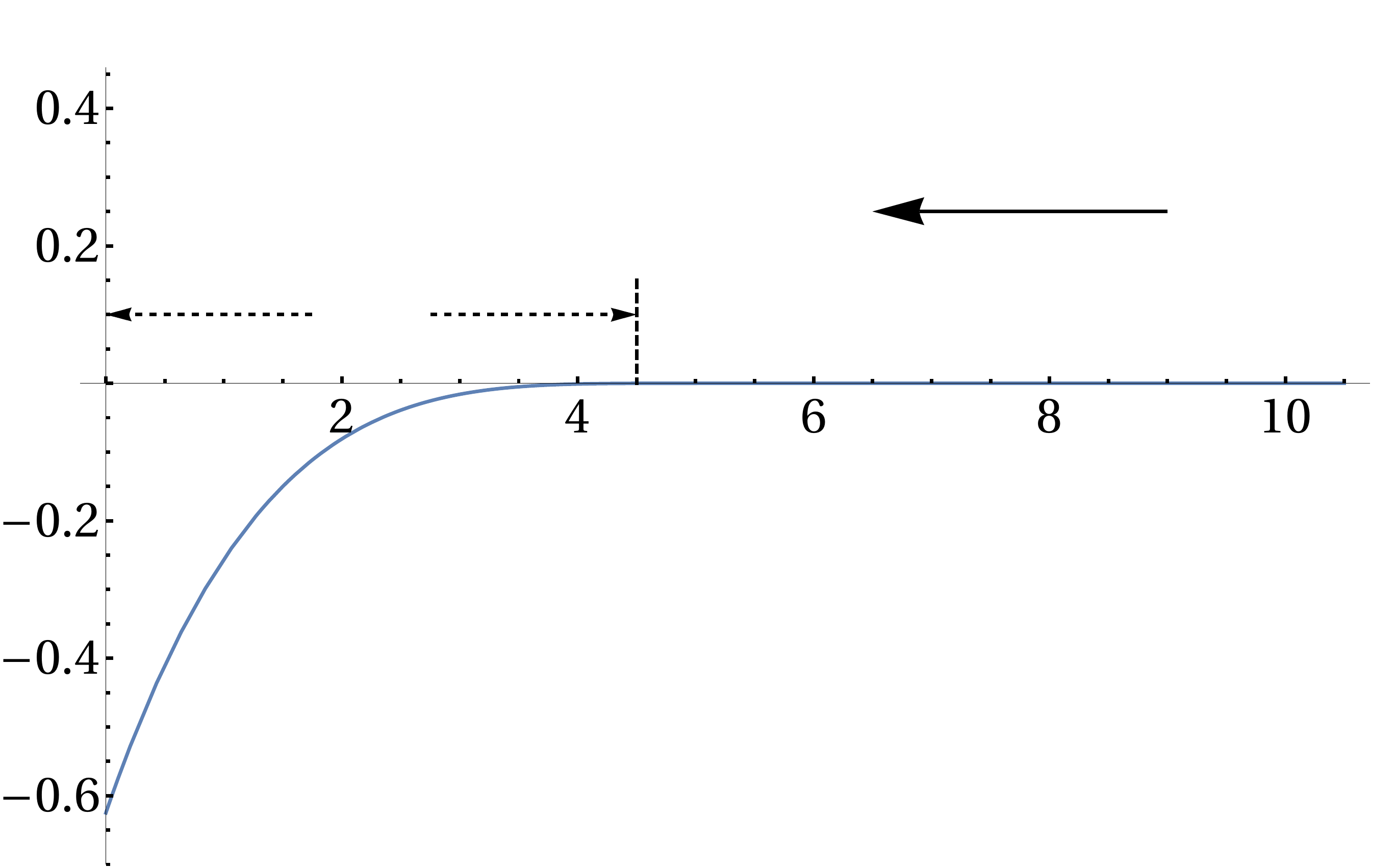} 
\setlength{\unitlength}{0.01\columnwidth}
\begin{picture}(0,0)(100,0)
\put(54,37){\scalebox{1.1}{$\psi=e^{-ikx}+\sqrt{R}e^{i\varphi} e^{ikx}$}}
\put(25,37){\scalebox{1.1}{$ x_R $}}
\put(97,32.5){\scalebox{1.1}{$x$}}
\put(5,56){\scalebox{1.0}{$|V(x)|$}}
\end{picture}
\caption{\label{scheme} 
The scheme used to characterize a CAP and determine its reflection coefficient. 
The edge of the grid is taken to be $x=0$, and we require $\psi(x$=$0) =0$ as is typical in solving the TDSE.
We assume incidence from the right as indicated. 
We define the absorption range $x_R$ from the distance at which the absorbing
potential decreases beyond a cutoff value $V_c$ and can be neglected,
$|V(x_R)| = V_c$.} 
\end{figure}
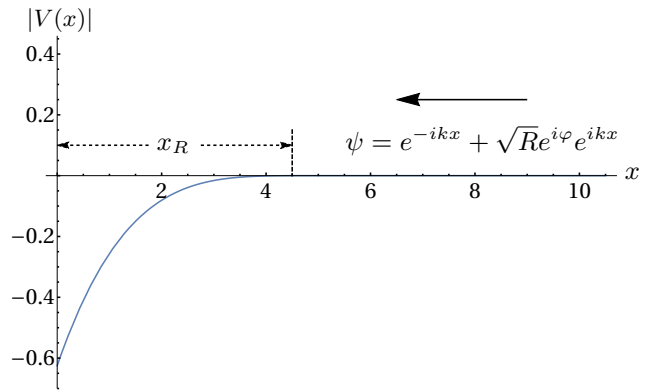

Since a time-dependent wavepacket can always be written as a superposition of the time-independent scattering states, 
we use the time-independent reflection coefficient as a quantitative tool for characterizing and designing 
an optimal CAP. 
We will require the CAP's reflection coefficient to remain  below a cutoff value $R_c$, $R \leqslant R_c$, over a
given energy range $E_{\rm min} \leqslant E \leqslant E_{\rm max}$. 
Since the spatial region devoted to the CAP near the edge of the grid is unphysical, we wish to minimize the computing resources it consumes as much as possible. 
Therefore, in this work, our optimization  efforts focus on reducing the absorption range $x_R$, 
as defined in Fig.~\ref{scheme}, while meeting the absorption criteria above.

\begin{table}  
\caption{The CAPs considered in this work, both from the literature and proposed in this work.} 
\begin{ruledtabular} 
\begin{tabular}{ll } 
 CAP type & $V(x)$ (units of $\hbar^2/2 m$) \rule{0pt}{2.4ex} \\ \hline
 quadratic~\cite{Riss_Meyer_JCP_1996,Muga_PhysRep_2004}       & $-i \alpha^2 (x-x_0)^2$   \rule{0pt}{2.6ex}  \\ 
 cosine masking function~\cite{Kulander_MaskingFunc_PRA_1992} & $-i \alpha^2 \log\!\left\{ \sec \left[(x_0-x)/ \beta \right] \right\} $  \\ 
 M-JWKB~\cite{Manolopoulos_JCP_2002}                     & $-i k_\text{min}^2 \epsilon(x)$ \\
 quartic~\cite{Riss_Meyer_JCP_1996,Muga_PhysRep_2004}  & $-i \alpha^2 (x-x_0)^4 $ \\ 
 pseudo-exponential~\cite{TISE_Balint-Kurti_JCP_1992}  & $ -i \alpha^2 e^{-\beta/(x_0-x)}  $ \\ 
 P\"oschl-Teller~\cite{KOSLOFF_JCompP_1986}            & $ -i \alpha^2 \text{sech}^2[(x+x_0)/{\beta}] $  \\ 
 single-exponential (present)                          & $- \alpha^2   e^{-x/\beta} $ \\ 
 double-exponential (present)                          & $ -\alpha_1^2 e^{-x/(2 \beta)} -i \alpha_2^2 e^{-x/\beta} $  \\ 
 double-sinh (present)                                 & $ -\alpha_1^2/(2\sinh[x/(2 \beta)])$\\
                                                       & \hfill$ -i \alpha_2^2/(2\sinh[x/(2\beta)])^2$ 
\end{tabular} 
\end{ruledtabular} 
\label{CAPs}
\end{table} 

We study one-dimensional CAPs since they can be easily adapted to higher dimensions, 
obtaining the reflection coefficient $R$ by solving the Schr\"odinger equation,
\begin{align} 
 \left[ -\frac{\hbar^2}{2 m} \frac{d^2}{d x^2} + V(x) \right] \psi = E \psi, 
\label{TISE} 
\end{align} 
as indicated schematically in Fig.~\ref{scheme}.  We consider the potential
$V(x)$ to be one of the CAPs listed in Table~\ref{CAPs}.  
The shapes of all the CAPs considered are generically as in Fig.~\ref{scheme}
and are controlled by the following parameters: $\alpha^2$ is the strength of the potential, 
$\beta$ mainly determines its width, and $x_0$ is a shift.  These are the parameters that
will be varied to optimize the CAPs.

The JWKB-based CAP obtained by Manolopoulos~\cite{Manolopoulos_JCP_2002}\,---\,labeled M-JWKB 
in Table~\ref{CAPs}\,---\,is qualitatively different from the others, however, in that it requires
no optimization.  This simplicity is certainly one of its strengths and derives from the fact that 
its reflection coefficient effectively decreases monotonically from unity at zero energy to $e^{-\sqrt{2}\pi/\delta}$
at infinite energy.  Its explicit expression is in terms of the Jacobi elliptic function $\text{cn}(u,k)$,
\begin{equation}
\epsilon(x)=\sqrt{\text{cn}^{-4}[2\delta k_\text{min}(x_0-x)/{\sqrt{2}},{1}/{\sqrt{2}}]-1},
\end{equation}
with $x_0$=$2.622/(2\delta k_\text{min})$ where
$k_\text{min}$=$\sqrt{2 m E_\text{min}/\hbar^2}$~\cite{Manolopoulos_JCP_2002}.  One simply chooses
$\delta$ from the condition $R(E_\text{min})=R_c$.

The first five CAPs in the table are defined to be non-zero only for $0\leqslant x\leqslant x_0$ and to vanish 
identically for $x> x_0$.  The remaining
CAPs are defined for all $x$, but vanish exponentially with $x$.  The first four CAPs are some of the most
commonly used, with the cosine masking function recast as a CAP using 
$e^{-i V(x) \Delta t} \sim \cos^{1/8}[(x-x_0)/\beta]$. 

We include the P\"oschl-Teller potential because it is well known to be reflectionless for 
specific {\em real} values of $i \alpha^2$, suggesting that it might have advantageous properties as a CAP.
It can be shown analytically, however, that this property no
longer holds for complex $i \alpha^2$. In the process of optimizing it for the present
purposes, we found that shifting its center off the grid 
minimized $x_R$, leaving only its exponential tail on the grid.  This result suggested using 
instead the simpler family of exponential CAPs included in the table.

We calculate the reflection coefficient numerically using the finite-element discrete-variable 
representation (FEDVR)~\cite{FEDVR_PRA_McCurdy_2001,FEDVR_PRE_Schneider_2006} and eigenchannel R-matrix 
method~\cite{RMatrix_Greene_RMP}. 
The reflection coefficient can also be calculated analytically for several of the potentials in Table~\ref{CAPs}.  
However, we give the analytic solutions (derivations in the appendices) only for the 
CAPs we propose\,---\,namely, the single-exponential and 
double-exponential CAPs.  The double-sinh potential has no analytic solution
to the best of our knowledge.  In these cases, we confirmed that the R-matrix 
reflection coefficients agreed with the analytical $R$ to several significant digits.

Since our primary goal is to systematically design an absorbing potential for  the strong-field 
ionization problem with predetermined properties, we will use atomic units for the remainder of our discussion.
Our results can be readily applied to other problems, though, using the derivations in the appendices
in which the masses and SI units are explicitly retained.

\section{\label{OptParam} Optimization of Proposed CAPs} 

We demonstrate our optimization procedure in detail below first for
the single-exponential CAP since it is the simplest to optimize.  It also establishes
a few key results important for the optimization of our recommended CAPs: the double-exponential and
double-sinh potentials.  Whether the solution is analytical or numerical, 
the procedure we describe for optimization is the same and can be applied to
other CAPs as well.  In fact, this is what we have done for the comparison in Sec.~\ref{CAPComp}.

The values of $R_c$, $E_\text{min}$, and $E_\text{max}$ that we will focus on for this discussion are
\begin{equation}
R_c=10^{-3},~  E_\text{min}=0.006\,\text{a.u.}, \text{ and }  E_\text{max}=3\,\text{a.u.}
\label{Params}
\end{equation}
We chose this energy range to cover 
$0.1 \hbar \omega  \leqslant E \leqslant 14U_p$ for an 800-nm laser pulse 
at $10^{14}\,\text{W/cm}^2$ ($U_p$ is the pondermotive energy: $U_p=I/4 \omega^2$ with
$I$ the intensity and $\omega$ the laser frequency).  This energy range includes 
essentially all photoelectrons one would expect to be produced in this typical pulse.
In momentum, which is more convenient for the analytical $R$, this range corresponds to 
\begin{equation}
k_\text{min}=0.110\,\text{a.u.} \text{ and } k_\text{max}=2.45\,\text{a.u.}
\label{MomRange}
\end{equation}
Note that 14$U_p$ exceeds the highest-energy electrons one would normally expect in a 
strong-field problem, but we will show below that this choice has little effect on the
resulting $x_R$. Finally, we use $V_c$=$10^{-4}\,\text{a.u.}$ to define $x_R$ from $|V(x_R)|$=$V_c$.
 
\subsection{Single-Exponential CAP} 
\label{seCAP}

We take the single-exponential CAP to have the form
\begin{equation}
V(x) =  -\frac{\hbar^2\alpha^2}{2m} e^{-x/\beta}
\end{equation}
Its reflection coefficient, as shown in App.~\ref{expAppend}, is
\begin{equation} 
  R  = e^{4K \arg\lambda^2} \left| \frac{J_{2 i K }(2 \lambda)}{J_{-2 i K }(2 \lambda)}   \right|^2, 
\label{seCAP_R}
\end{equation} 
where the unitless momentum is $K = k\beta$ with $k = \sqrt{2 E} $ and the unitless
potential strength is $\lambda = \alpha  \beta $. 
To achieve our goal of minimizing $x_R$, we must find the optimal $\lambda$ and $\beta$.  

\subsubsection{Purely imaginary potential}

\begin{figure} 
\includegraphics[width=0.9\columnwidth]{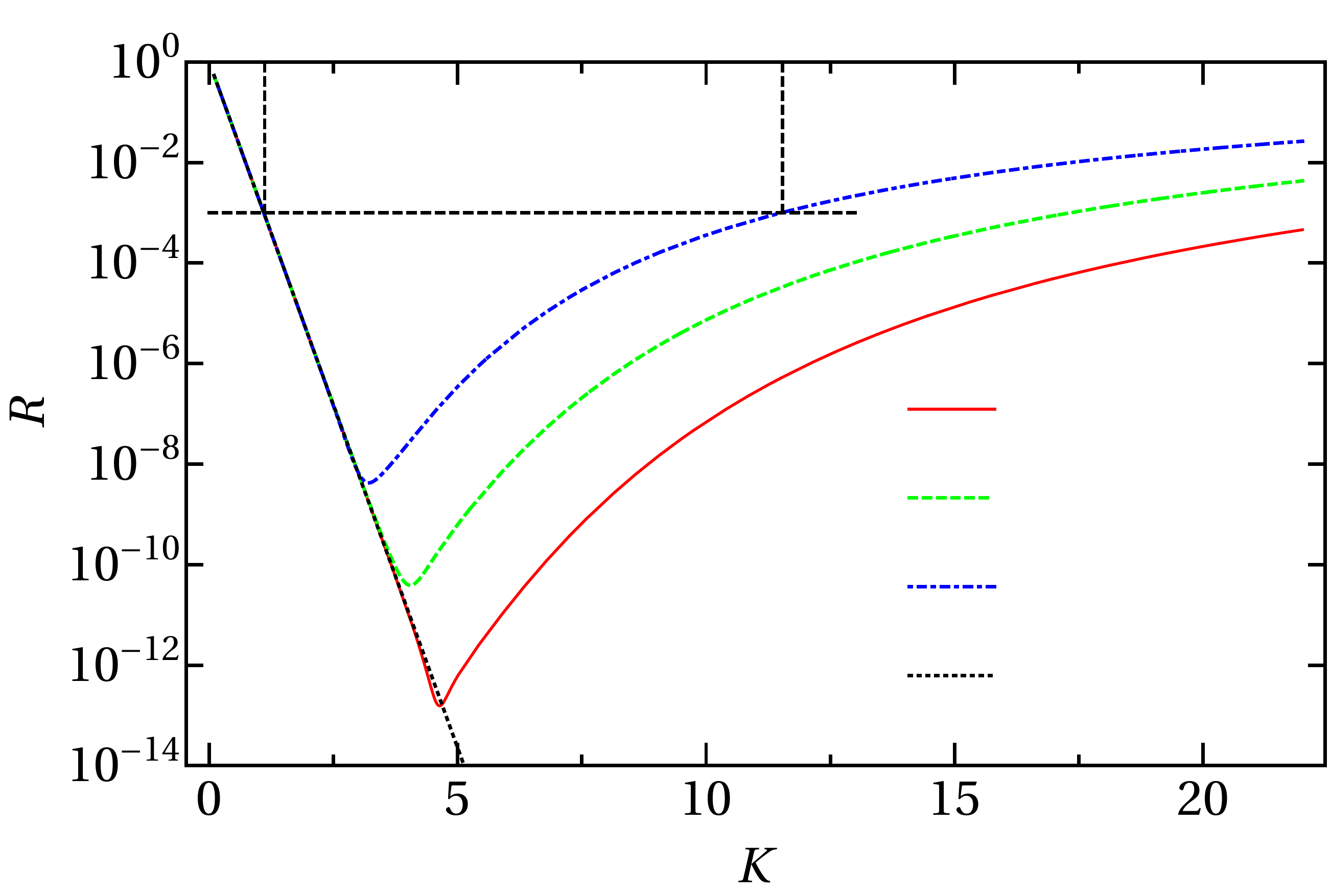}
\setlength{\unitlength}{0.01\columnwidth}
\begin{picture}(0,0)(100,0)
\put(77,32){\scalebox{1.0}{$\lambda^2=84.7i$}}
\put(77,26){\scalebox{1.0}{$\lambda^2=60i$}}
\put(77,20){\scalebox{1.0}{$\lambda^2=40i$}}
\put(77,14){\scalebox{1.0}{$e^{-2\pi K}$}}
\put(30,42){\scalebox{0.9}{$R_c=10^{-3}$}}
\put(28,51){\scalebox{0.9}{$K_{\rm min}$}}
\put(62,51){\scalebox{0.9}{$K_{\rm max}$}}
\end{picture}
\caption{\label{expRef} 
Examples of the reflection coefficient $R(K)$ for a single-exponential CAP with different potential strengths.
The predicted small-$K$ behavior, $e^{-2\pi K}$, is shown for comparison.
The unitless limits $K_{\rm min}$ and  $K_{\rm max}$ for which 
$R(K)\leqslant R_c$ holds are also indicated.
} 
\end{figure}
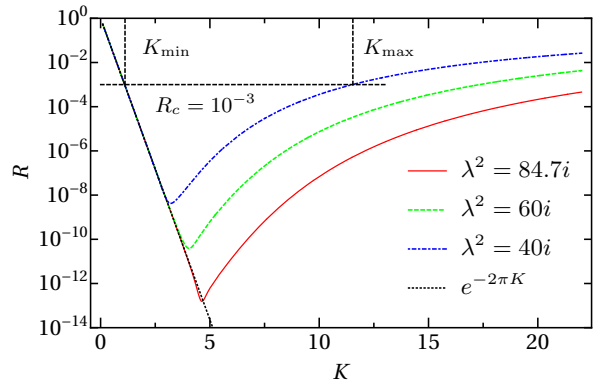

For a purely imaginary potential, $\lambda^2 \propto i$, Fig.~\ref{expRef} shows the behavior of $R$ 
as a function of $K$.  As the figure suggests, one can show from Eq.~(\ref{seCAP_R}) that
\begin{equation}
R\xrightarrow[K\rightarrow 0]{} e^{-2 \pi K}.
\label{seCAP_R_smallK}
\end{equation}
As can also be seen in the figure, 
increasing the strength $|\lambda^2|$ of the CAP means this exponential decrease continues to 
larger $K$ and the large-$K$ tail decreases. 

We can thus use Eq.~(\ref{seCAP_R_smallK}) to write
\begin{equation}
K_\text{min}=-\frac{1}{2\pi}\log R_c,
\end{equation}
giving
\begin{equation*} 
 K_\text{min} = 1.10
\end{equation*} 
for $R_c=10^{-3}$.
From $K_{\rm min} = k _{\rm min} \beta$, the scale $\beta$ is therefore determined: 
\begin{equation*} 
 \beta = \frac{K_{\rm min}}{k_{\rm min}} = 10.0\,\text{a.u.}. 
\end{equation*} 
We can now find the required $\lambda^2$ from 
\begin{equation}
R(K_\text{max})=R\biggl(\frac{k_\text{max}}{k_\text{min}} K_\text{min}\biggr)=R_c
\label{lambdaEqn}
\end{equation}
since $K_{\rm max}$=$k_{\rm max} \beta$. 
Solving this equation gives 
\begin{align} 
\lambda^2 = 84.7 i \text{ and }
 x_R = 83.4\,{\rm a.u.} 
\end{align} 

This example illustrates the fact that $x_R$ can only be substantially decreased if $\beta$ is decreased.  Thus, 
$K_{\rm min}$ and $k_{\rm min}$ determine $x_R$, and $K_{\rm min}$ is set by the form of the CAP and its parameters. 
In general, the smaller $k_{\rm min}$ is, the more difficult absorption becomes. 
Roughly speaking,  this behavior can be traced to  the need for  $x_R$ 
to be large enough for the potential to contain the longest wavelength to be absorbed. 

\subsubsection{Complex potential}
\label{ComplexPot}

Given $k_{\rm min}$, decreasing $\beta$ further requires decreasing $K_{\rm min}$. 
This is not possible with a purely imaginary  single-exponential CAP, so we must allow $\lambda^2$ to be complex. 

The reflection coefficient in Eq.~(\ref{seCAP_R}) still holds for complex $\lambda^2$ and looks generically like 
those displayed in Fig.~\ref{expRef}\,---\,with the exception that 
\begin{equation}
R\xrightarrow[K\rightarrow 0]{}  e^{-4\pi K+4K\arg\lambda^2}.
\label{seCompSmK}
\end{equation}
This small-$K$ behavior suggests that the best way to reduce $K_\text{min}$\,---\,and thus $\beta$ and $x_R$\,---\,is to 
make $\arg\lambda^2$ small (since $\arg\lambda^2$ must be positive to have absorption).  That is, we should
make $\text{Re}\,\lambda^2$ much larger than $\text{Im}\,\lambda^2$.  The fastest decay
one can achieve with this approach is $e^{-4\pi K}$ which, in turn, sets the limit on how small $K_\text{min}$
can be.

The physical origin of this faster low-$K$ decrease is clear: the real part of the potential is attractive and 
accelerates the wave before it encounters the imaginary part of the potential~\cite{Muga_PhysRep_2004}.  Absorption thus 
occurs at a shorter wavelength  where absorption can be efficient with a much smaller $x_R$.
Since $\text{Im}\,\lambda^2$ must be large enough for sufficient absorption, however, $\arg\lambda^2$
cannot be zero.  The optimum value will be a compromise between these two factors.

\begin{figure} 
\includegraphics[width=\columnwidth]{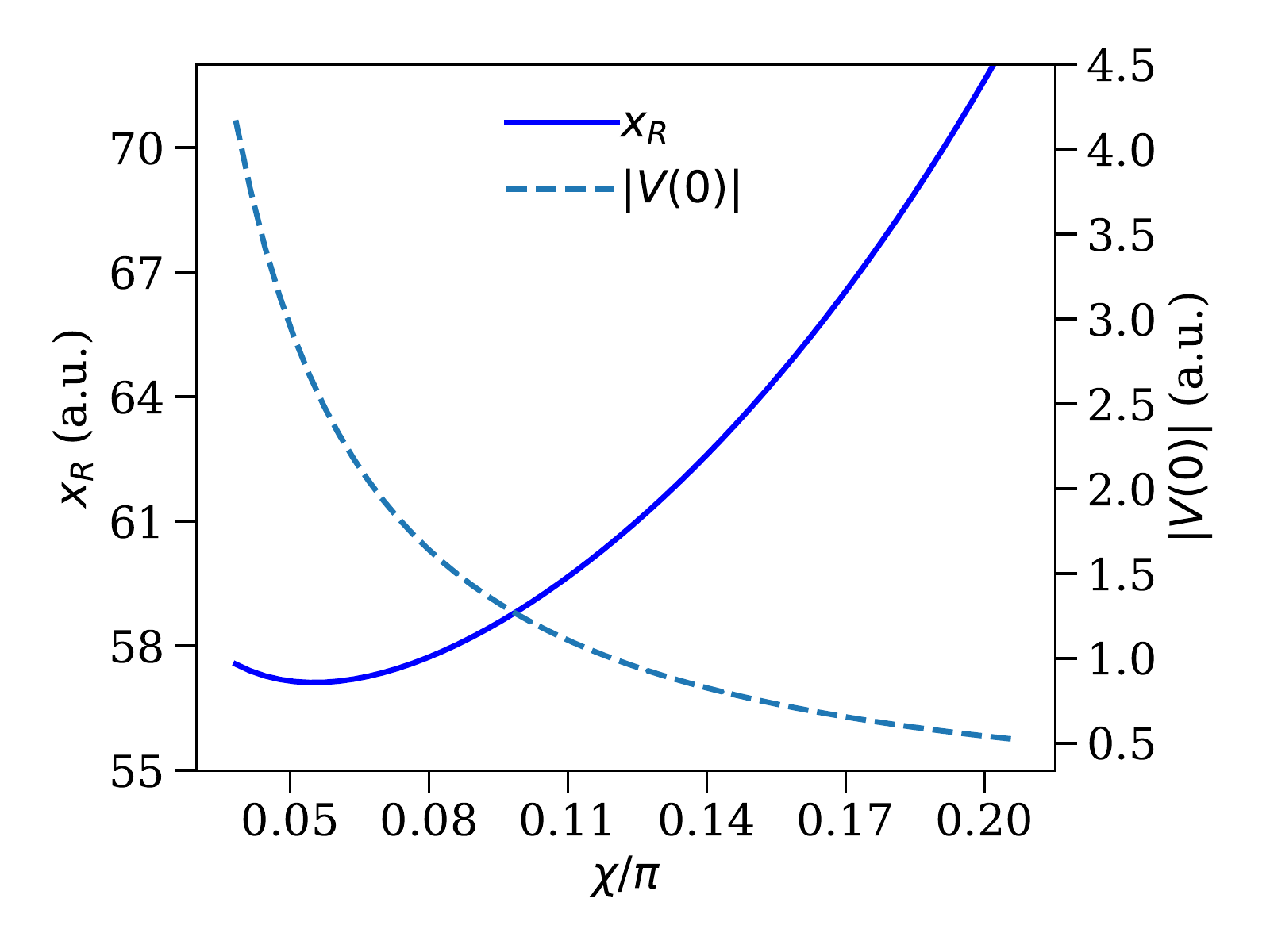}
\caption{\label{seOpt} 
Absorption range and potential depth
for a single-exponential CAP, showing their dependence 
on the complex phase of $\lambda$.  The magnitude of $\lambda$ is
determined at each $\chi$ by solving Eq.~(\ref{lambdaEqn}).
} 
\end{figure}

To determine the magnitude of the improvement in $x_R$, we use $\lambda=|\lambda|e^{i\chi}$ and
the small-$K$ behavior in Eq.~(\ref{seCompSmK}) to write
\begin{equation}
K_{\rm min}=\frac{\log R_c}{4 (2\chi-\pi)}.
\end{equation}
From this, we can find $\beta$ and $K_{\rm max}$ for a given $\chi$. Combining everything
and simplifying reduces the problem to solving Eq.~(\ref{lambdaEqn})
for $|\lambda|$ with $R(K)$ from Eq.~(\ref{seCAP_R}).  The resulting 
$x_R$ as a function of $\chi$ is shown in Fig.~\ref{seOpt}.  

The figure shows that the optimization problem has been reduced to a one-dimensional
minimization of $x_R$ with respect to $\chi$.  As expected, the solution, 
\begin{equation}
x_R=57.1\,\text{a.u.}
\end{equation}
at $\chi$=0.055$\pi$ (with $|\lambda^2|=165$),
lies at small $\chi$.  Adding a real part to the absorbing potential has thus reduced
the absorption range by 32\% over the purely imaginary single-exponential CAP.

Figure~\ref{seOpt} also shows that at the
optimal $x_R$, the potential is 2.62\,a.u. deep.  This is roughly equal to $E_\text{max}$,
leading to local kinetic energies of approximately $2E_\text{max}$ and thus
requiring a much denser spatial grid in the absorption region.
Guided by the figure, however, we see that 
a modest few-percent increase in $x_R$ to 58.9\,a.u. ($\lambda^2=97e^{i \pi/5}$) reduces $|V(0)|$ 
to 1.25\,a.u., making it more computationally attractive.  Further reduction
in $|V(0)|$ can, of course, be achieved\,---\,at the expense of $x_R$.

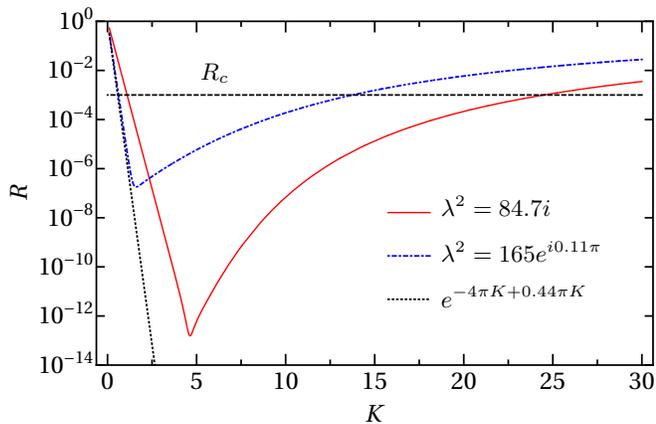
\begin{figure} 
\includegraphics[width=\columnwidth]{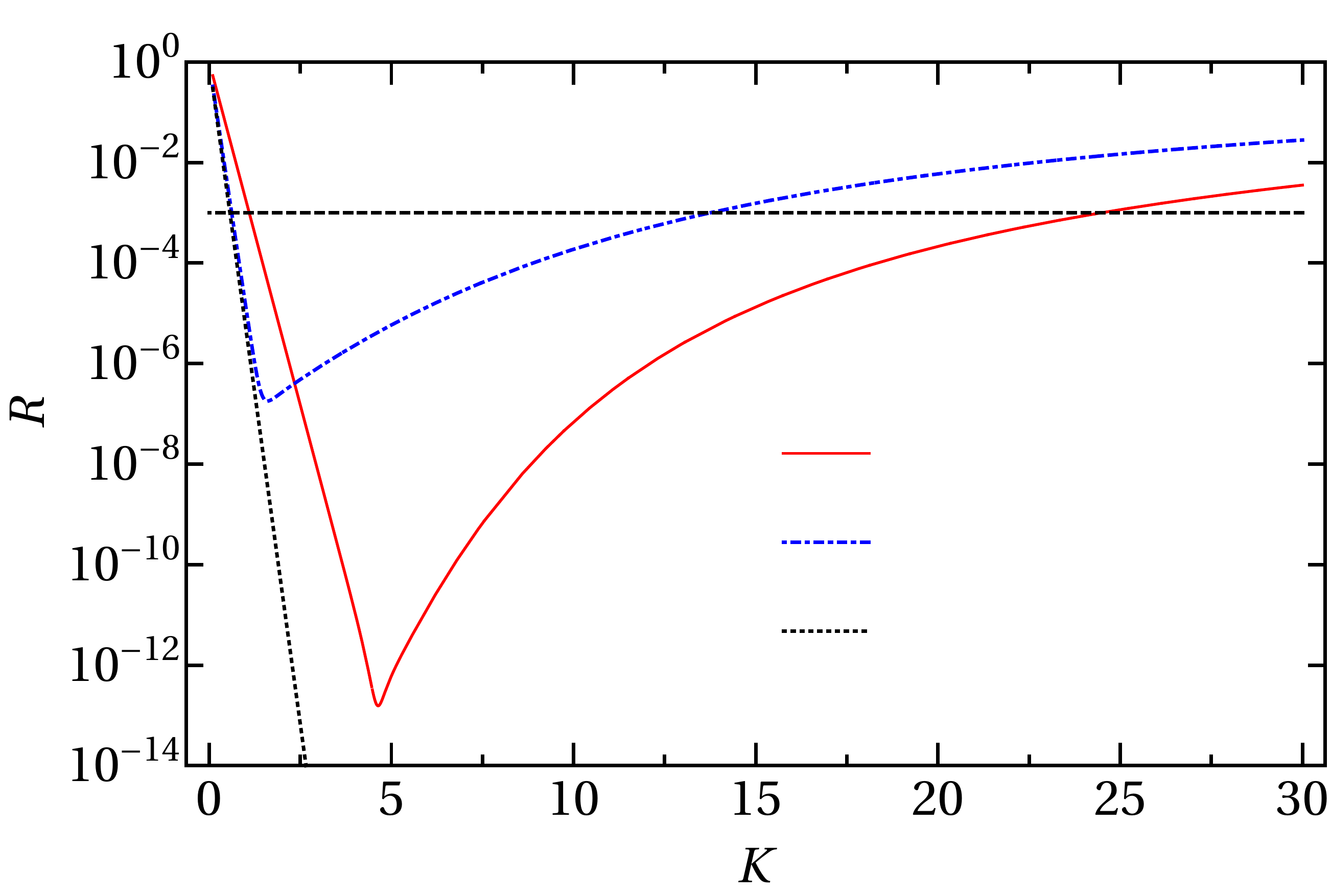}
\setlength{\unitlength}{0.01\columnwidth}
\begin{picture}(0,0)(50,0)
\put(67,37){\scalebox{1.0}{$\lambda^2=84.7i$}}
\put(67,30){\scalebox{1.0}{$\lambda^2=165 e^{i0.11\pi}$}}
\put(67,23){\scalebox{1.0}{$e^{-4\pi K+0.44\pi K}$}}
\put(30,58){\scalebox{1.0}{$R_c$}}  
\end{picture}
\caption{
The optimum reflection coefficients for purely imaginary and complex single-exponential
CAPs as a function of the unitless momentum.} 
\label{seOptR} 
\end{figure}

Figure~\ref{seOptR} shows the optimum $R$ for both the purely imaginary single-exponential
CAP of the previous section and the complex single-exponential CAP of the present section.
The coefficients satisfy $R\leqslant R_c$ for different ranges of the scaled momentum  $K$ but
the same range of the physical momentum  $k$.  The range of $K$ covered by the complex
CAP is smaller than for the imaginary CAP by the ratio of their respective $K_\text{min}$'s.

\subsection{\label{deCAP}Double-Exponential CAP} 

It has long been known, of course, that adding a real potential improves CAP performance~\cite{Muga_PhysRep_2004}.
And, given the improvement to the single-exponential CAP afforded by doing so, it is natural
to ask whether we can do even better with a more flexible complex potential.  

Since we want to retain the  ability to calculate $R$ analytically and since the real part must have longer range than
the imaginary part, we choose the CAP to be
\begin{align} 
 V(x) = -\frac{\hbar^2\alpha_1^2}{2m} e^{-{x}/{2 \beta}} -i  \frac{\hbar^2\alpha_2^2}{2m} e^{-{x}/{\beta}}.
\end{align}  
The reflection coefficient, as shown in App.~\ref{doubleExpAppend}, can be written in terms of the 
confluent hypergeometric function as
\begin{align} 
  R = \left| \frac{_1F_1(\eta+2iK,1+4iK,-4\gamma^3 \lambda_2 )}{_1F_1(\eta-2iK, 1-4iK, -4 \gamma^3 \lambda_2)} \right|^2 
  \label{deR}
\end{align} 
where 
\begin{align*} 
  K &= k \beta            & \lambda_1 &= \alpha_1 \beta                     & \lambda_2 &= \alpha_2 \beta \\
  \gamma&= e^{i\pi/4}  & \varLambda &= \frac{\lambda_1^2 }{ \lambda_2} &  \eta &= \frac{1}{2} - {\gamma \varLambda}      .
\end{align*} 

\subsubsection{$\lambda_1$ and $\lambda_2$ independent}

Given the extra potential parameter, optimizing the double-exponential CAP is clearly more challenging than
for the single-exponential CAP.  And, the complicated expression for $R$ only exacerbates the task.  It would
therefore be convenient to find a regime in which $\lambda_1$ and $\lambda_2$ are independent since this
would greatly simplify the optimization.  

\begin{figure} 
\includegraphics[width=\columnwidth]{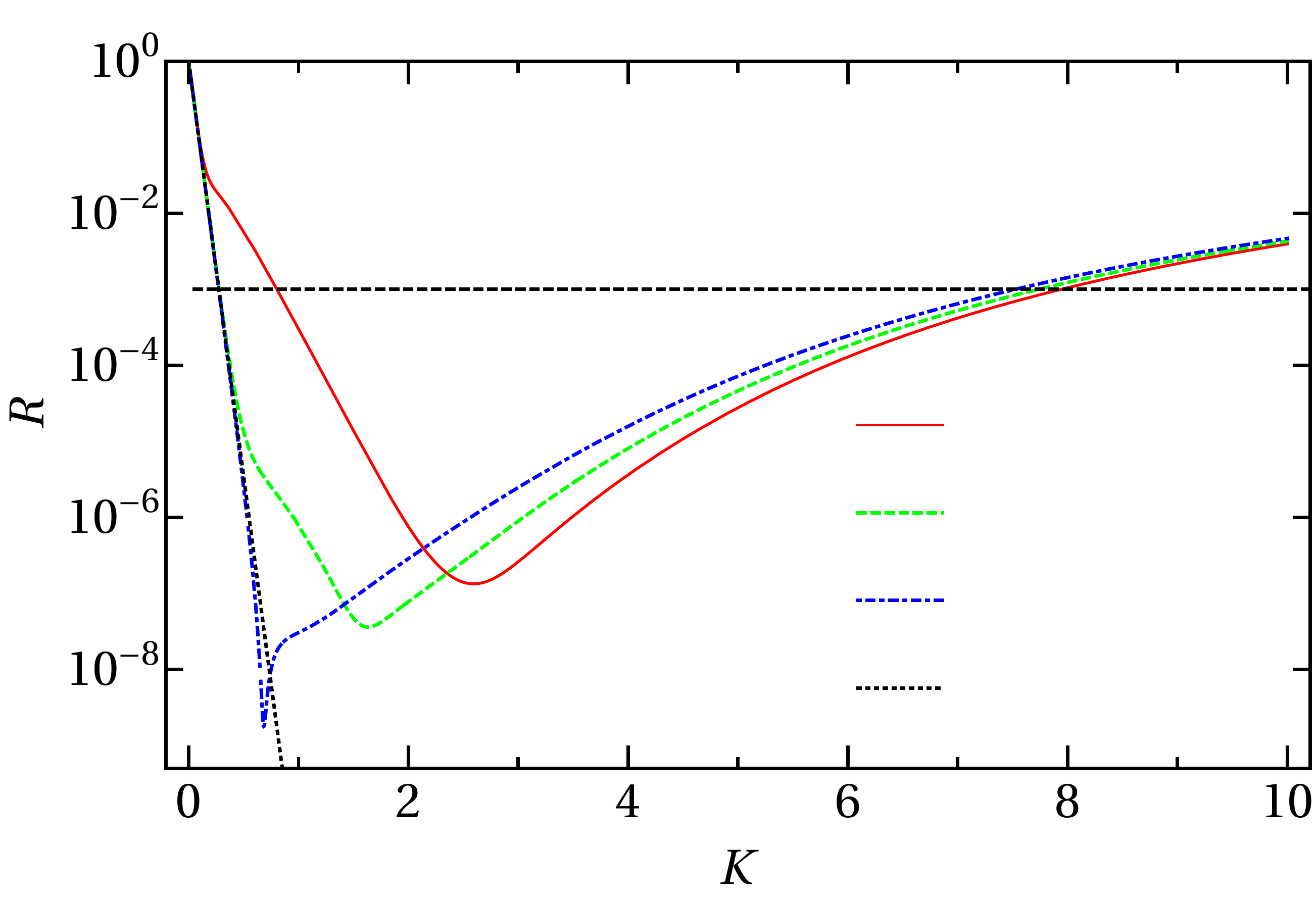}
\includegraphics[width=\columnwidth]{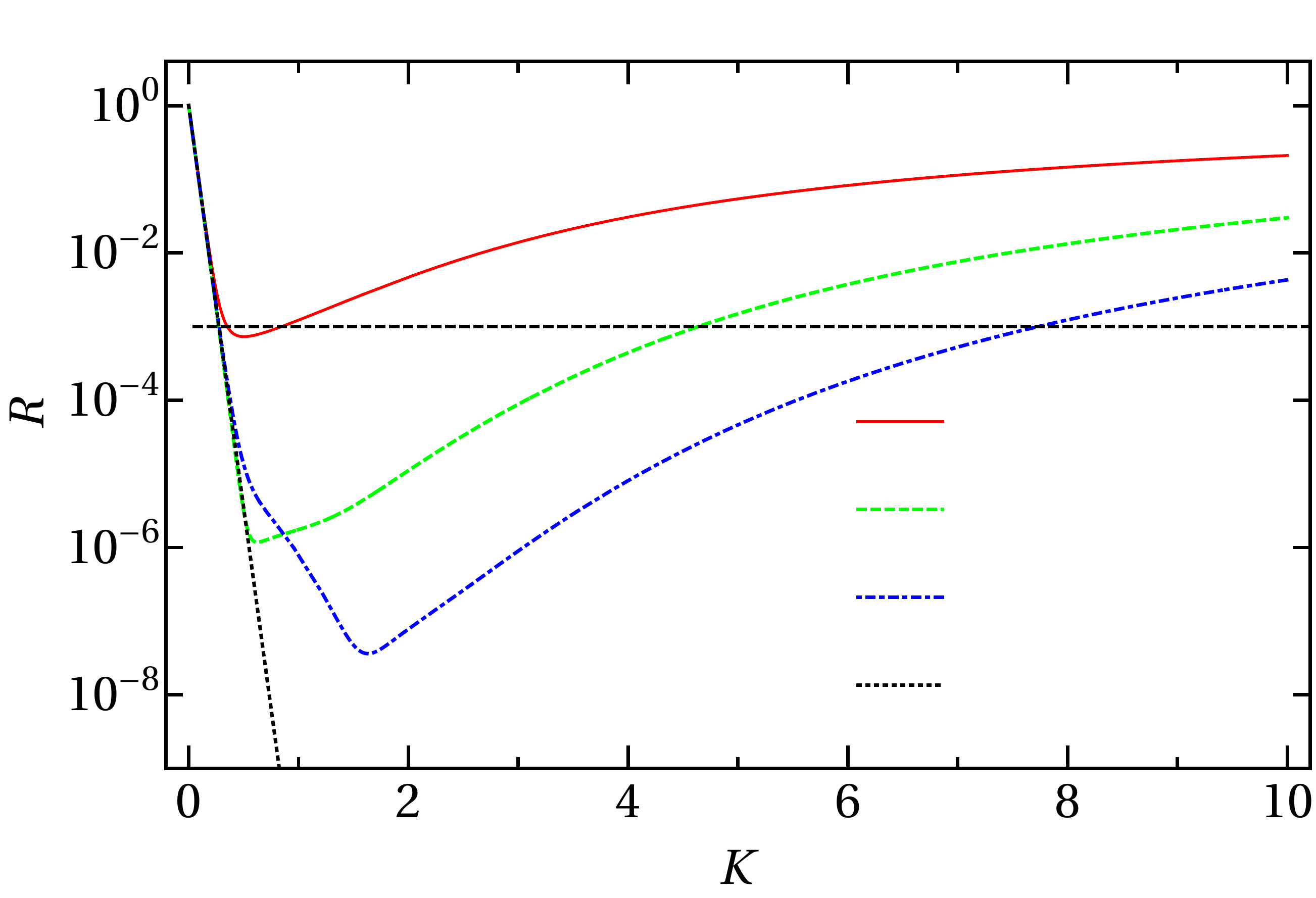}
\setlength{\unitlength}{0.01\columnwidth}
\begin{picture}(0,0)(50,0)
\put(92,131){\scalebox{1.0}{(a)}}
\put(74,108.2){\scalebox{1.0}{$\lambda_1^2=2$}}
\put(74,101.4){\scalebox{1.0}{$\lambda_1^2=7$}}
\put(74,94.6){\scalebox{1.0}{$\lambda_1^2=12$}}
\put(74,87.8){\scalebox{1.0}{$e^{-8\pi K}$}}
\put(35,122){\scalebox{1.0}{$R_c$}}  
\put(92,63){\scalebox{1.0}{(b)}}
\put(74,39.6){\scalebox{1.0}{$\lambda_2^2=8$}}
\put(74,32.8){\scalebox{1.0}{$\lambda_2^2=18$}}
\put(74,26){\scalebox{1.0}{$\lambda_2^2=28$}}
\put(74,19.2){\scalebox{1.0}{$e^{-8\pi K}$}} 
\put(35,49){\scalebox{1.0}{$R_c$}}  
\end{picture}
\caption{Illustration of the dependence of $R$ for a double-exponential CAP
on the potential strength:
(a) $\lambda_1$ dependence for $\lambda_2^2=28$, and  
(b) $\lambda_2$ dependence for $\lambda_1^2=7$.
}
\label{de_lambdaDep}
\end{figure}
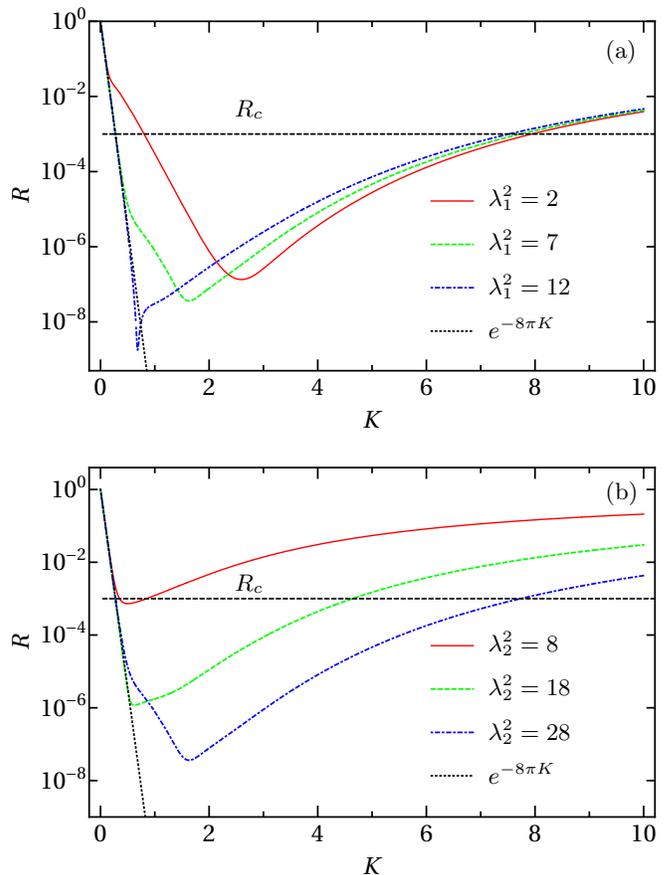

To this end, we show in Fig.~\ref{de_lambdaDep} the dependence of $R$ on $\lambda_1$ and $\lambda_2$.
Generally speaking, $\lambda_1$\,---\,the coefficient of the longer-ranged, real part of $V$\,---\,controls the low-energy
behavior, while $\lambda_2$\,---\,the coefficient of the shorter-ranged, imaginary part of $V$\,---\,controls the high-energy
behavior.  The underlying physical reasons for these connections are the same as discussed for the single-exponential
CAP.  

Figure~\ref{de_lambdaDep} also shows that for $\lambda_1$ and $\lambda_2$ large enough,
\begin{equation}
R\xrightarrow{}  e^{-8\pi K}
\label{deSmK}
\end{equation}
for $R\sim R_c$.  This behavior immediately shows the benefit of the double exponential since 
it falls faster than is possible with a single exponential, Eq.~(\ref{seCompSmK}), leading to 
a smaller $K_\text{min}$ and thus a smaller $x_R$. 

In the regime that Eq.~(\ref{deSmK}) holds, 
$K_\text{min}$ is independent of $\lambda_1$ and $\lambda_2$ and takes the value
\begin{equation}
K_\text{min}=-\frac{1}{8\pi}\log R_c.
\end{equation}
For $R_c$=$10^{-3}$, $K_\text{min}$=$0.275$ which is indeed much smaller than was possible with the 
single-exponential CAP.

Minimizing $x_R$ now requires fixing $\lambda_1$ to a large enough value that
Eq.~(\ref{deSmK}) holds ($\lambda_1^2\gtrsim 6$ is typically sufficient) and solving Eq.~(\ref{lambdaEqn})
for $\lambda_2$.  Using $R$ from Eq.~(\ref{deR}) and $K_\text{max}$=6.125, we find, for instance,
\begin{equation} 
 \lambda_1^2 = 6 \text{ and }\lambda_2^2 = 22.6, \text{ leading to } x_R = 42.4\,\text{a.u.}
\end{equation} 
and the reflection coefficient shown  in Fig.~\ref{de_Opt}.
There are, however, any number of combinations of $\lambda_1$ and $\lambda_2$ that satisfy Eq.~(\ref{lambdaEqn}).
Since $x_R$ for the double-exponential CAP is determined to a very good approximation by $\lambda_1$ alone,
though, one would typically choose the smallest possible $\lambda_1$ to obtain the smallest possible $x_R$.
At the same time, it should be noted that $x_R\propto\log\lambda_1$, so it is not terribly sensitive to
small changes in $\lambda_1$.
Choosing the smallest $\lambda_1$, however, also ensures that $|V(0)|$ is minimized, 
thereby keeping the numerical cost down.

\subsubsection{$\lambda_1$ and $\lambda_2$ not independent}

Although $x_R$=42.4\,a.u. is a significant improvement over the single-exponential result, $x_R$=57.1\,a.u., we can do better. 
The way to do this is to consider smaller $\lambda_1^2$
where there are particular combinations of $\lambda_1$ and $\lambda_2$ for which
$R$ falls off faster than Eq.~(\ref{deSmK}).  Such behavior permits smaller $K_\text{min}$ and
thus smaller $x_R$.
Of course, $\lambda_1$ and $\lambda_2$ are no longer independent in this regime,  but
it is still true that $\lambda_1$ largely\,---\,but not as completely as above\,---\,controls 
$K_\text{min}$ and $\lambda_2$, $K_\text{max}$. 

\begin{figure} 
\includegraphics[width=\columnwidth]{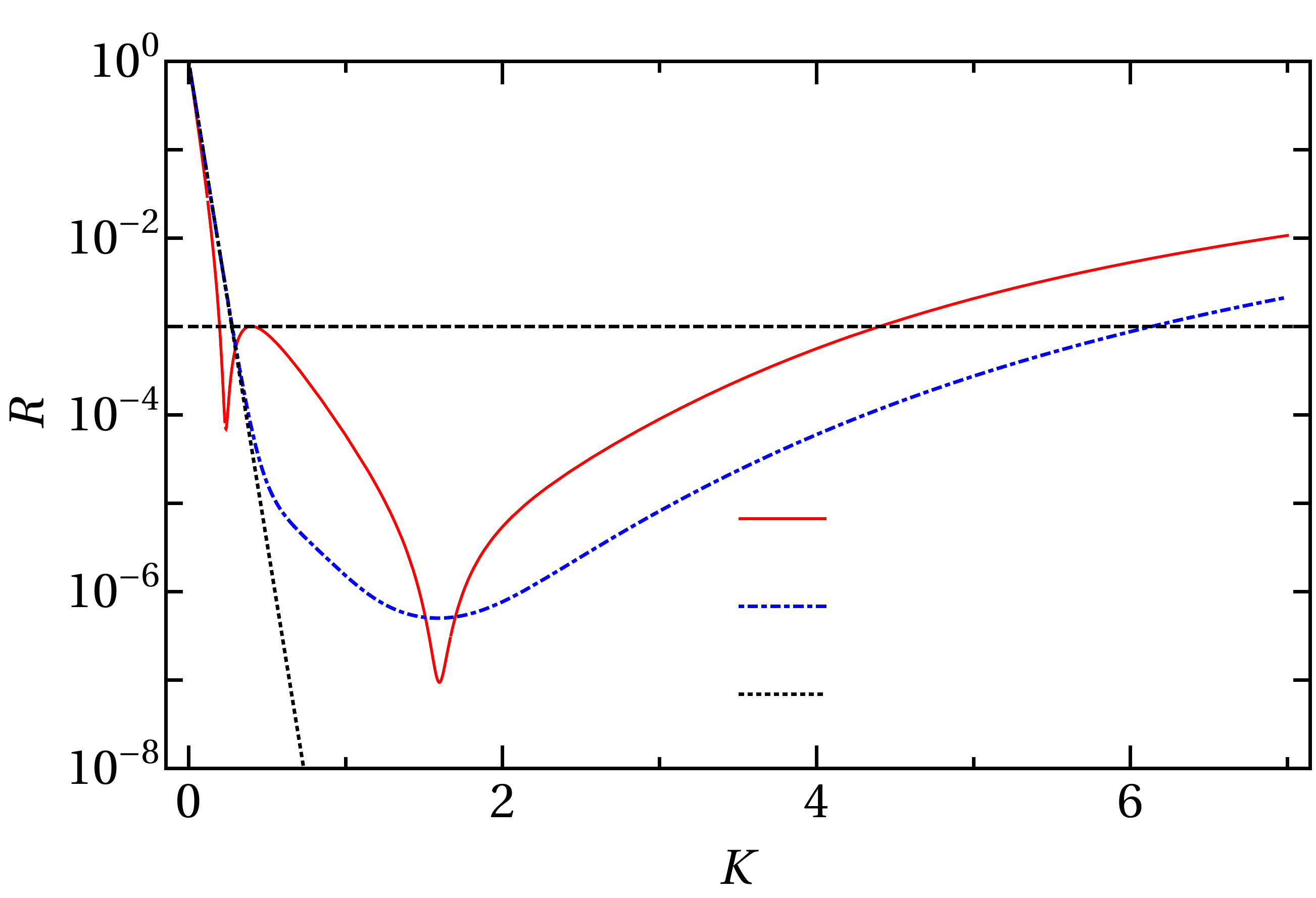}
\setlength{\unitlength}{0.01\columnwidth}
\begin{picture}(0,0)(50,0)
\put(64,32.5){\scalebox{1.0}{$\lambda_1^2=2.69, \lambda_2^2=16.3$}}
\put(64,25.5){\scalebox{1.0}{$\lambda_1^2=6, \lambda_2^2=22.6$}} 
\put(64,18.5){\scalebox{1.0}{$e^{-8\pi K}$}}
\put(35,50){\scalebox{1.0}{$R_c$}}  
\end{picture}
\caption{
The optimum reflection coefficient as a function of the unitless momentum
for the double-exponential CAP with $\lambda_1$ and $\lambda_2$ both 
independent and not independent. 
}
\label{de_Opt}
\end{figure}
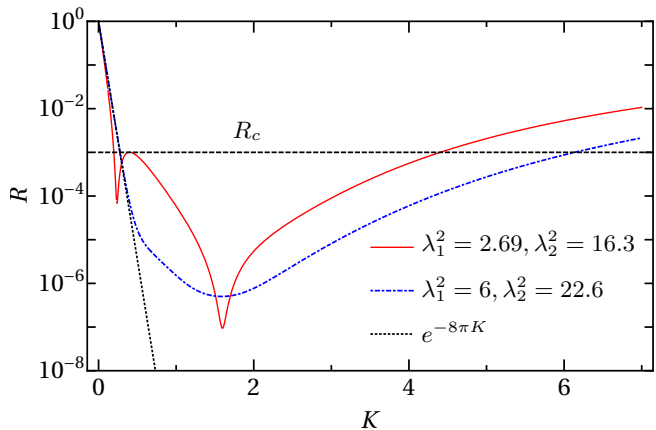 

Figure~\ref{de_Opt} illustrates the small-$K$ behavior that we want to take advantage of.  For this
combination of $\lambda_1$ and $\lambda_2$, $R$ dips below the exponential from Eq.~(\ref{deSmK})
for $R\sim R_c$ as seen in the figure.  At this and other such parameter combinations, a local minimum develops in $R$ 
at $K$ near $K_\text{min}$ as shown in the figure.  
In practice, one searches for these $\lambda_i$ combinations to minimize $K_\text{min}$ while
simultaneously ensuring that the local maximum in $R$ remains below $R_c$.  

To find the minimum value of $K_{\rm min}$, we take advantage of its weak
dependence on $\lambda_2$ by first minimizing with respect to $\lambda_1$ for some reasonable choice of $\lambda_2$. 
With this value of $\lambda_1$, we then solve Eq.~(\ref{lambdaEqn})
for $\lambda_2$.  Since there is a weak dependence on $\lambda_2$, though, $K_\text{min}$ must be re-minimized 
for $\lambda_1$ with this new $\lambda_2$. Then, Eq.~(\ref{lambdaEqn}) must again be solved and
the iteration continued until sufficient convergence in $\lambda_1$ and $\lambda_2$ is obtained.
Typically, only a handful of iterations are necessary to find 3 digits.  More sophisticated methods
of performing the constrained minimization of $x_R(\lambda_1,\lambda_2)$ could, of course, be employed as well.

As above, there are many combinations of $\lambda_1$ and $\lambda_2$ that give the smallest $K_\text{min}$,
$K_\text{min}=0.197$.  But, our ultimate goal of minimizing $x_R$ leads us to choose the smallest $\lambda_1$ possible. 
One convenient example for the optimal values is
\begin{align} 
  \lambda_1^2 = 2.69 \text{ and } \lambda_2^2 = 16.3,   \nonumber  
\end{align} 
which leads to 
\begin{align} 
 \beta = 1.79\,\text{a.u.} \text{ and } x_R =29.9\,\text{a.u.}. \nonumber 
\end{align} 
The corresponding $R$ is shown in Fig.~\ref{de_Opt}.
Although difficult to prove, this choice appears to be the global optimum for this
choice of $E_\text{min}$, $E_\text{max}$, and $R_c$.

\subsection{Double-sinh CAP}
\label{dsCAP}

While straightforward, the optimization procedure outlined above for
achieving such a substantial reduction in $x_R$ is somewhat tedious.
Fortunately, it needs to be done only once for a given $R_c$ and ratio $E_\text{max}/E_\text{min}$.
Should one wish to change $R_c$ or only one of the energy limits, however, re-optimization
is required.  It turns out, though, that the latter limitation can be lifted without
compromising on $x_R$.

In general, one expects the reflection coefficient to be unity for
$E=0$ and $E\rightarrow\infty$, and this is the behavior displayed by all the reflection
coefficients we have shown.  Consequently, the reflection coefficient necessarily
satisfies $R(E)=R_c$ at both low and high energies.
As mentioned in Sec.~\ref{Method}, however, the M-JWKB CAP~\cite{Manolopoulos_JCP_2002}
produces an $R$ that decreases more-or-less monotonically to a value controllably less than unity
at infinite energy.  Its parameters thus do not depend on $E_\text{max}$, removing
the need for re-optimizing with changes in either $E_\text{min}$ or $E_\text{max}$.  Unfortunately,
$x_R$ for the M-JWKB CAP turns out to be 118\,a.u. because its $R$ falls off relatively
slowly at low energies, leading to a large $K_\text{min}$.

To retain both the small $x_R$ found for the double-exponential CAP and the 
advantageous high-energy behavior of the M-JWKB CAP, we have designed the 
double-sinh CAP:
\begin{equation}
V(x) = -\frac{\hbar^2}{2m} \frac{\alpha_1^2}{2 \sinh\!\frac{x}{2 \beta}}
      -i\frac{\hbar^2}{2m} \frac{\alpha_2^2}{4 \sinh^2\!\frac{x}{2 \beta}}.
\end{equation}
At large distances, this CAP reduces exactly to the double-exponential CAP, thus
possessing its nice long-wavelength, low-energy properties. At short distances,
this CAP is dominated by the $-i\alpha_2^2/x^2$ divergence of the second term.
It is this behavior that is inspired by the M-JWKB CAP and that leads to 
similarly desirable high-energy behavior.

\begin{figure} 
\includegraphics[width=0.9\columnwidth]{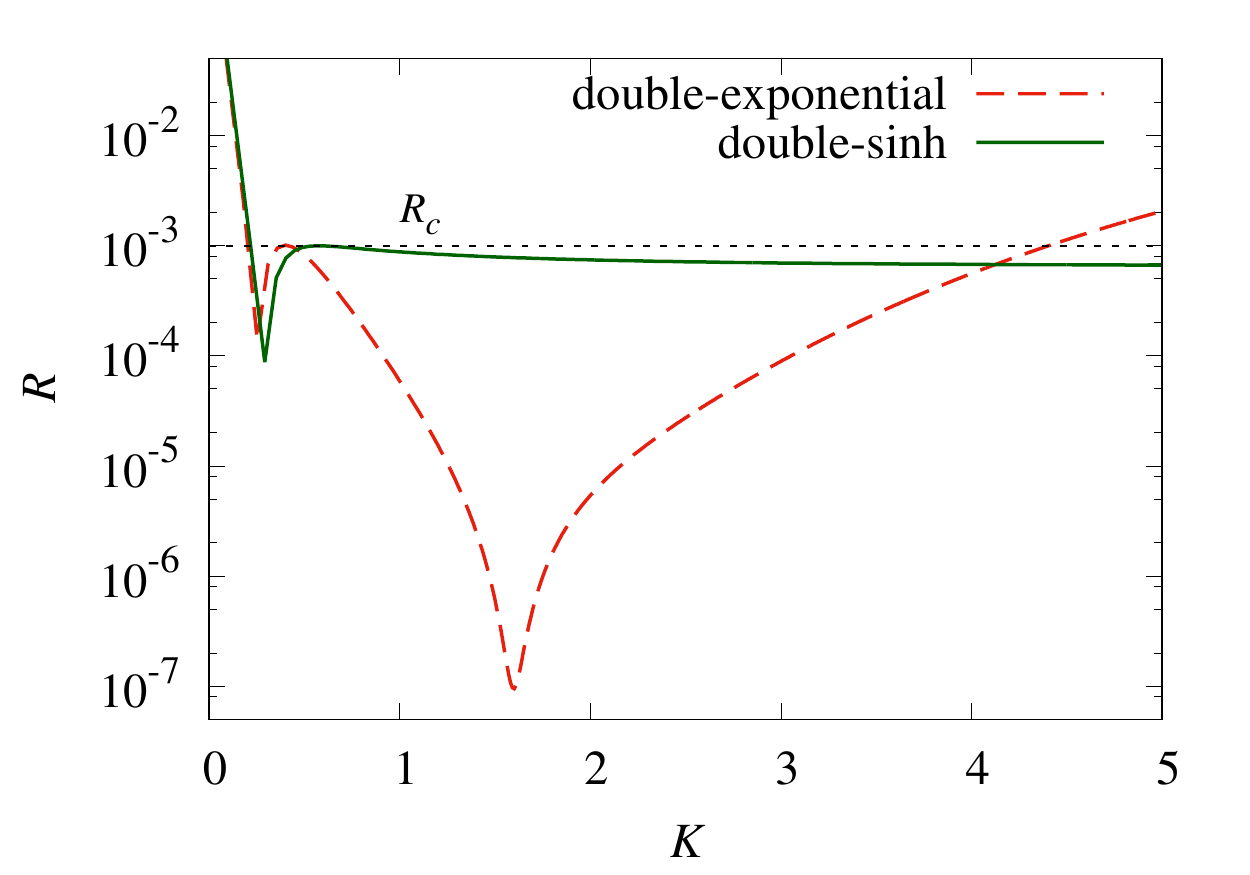}
\caption{Optimal $R$ for the double-sinh CAP along with the optimal $R$ 
for the double-exponential CAP for comparison using parameters in Table~\ref{optParams}, 
both as a function of the unitless momentum.
} 
\label{ds}
\end{figure}

Unlike the single- and double-exponential CAPs, $R$ for the double-sinh potential
is not analytic as far as we know (unless $\alpha_1$=0\,---\,in which case, it reduces
to one-half of the generalized P\"oschl-Teller potential~\cite{PT_1933}).  We must thus
calculate $R$ numerically, and the optimal result is shown in  Fig.~\ref{ds}
along with the optimal double-exponential result for comparison.  Their absorption
ranges are $x_R$=28.8\,a.u. and $x_R$=29.9\,a.u., respectively, confirming that
there is no compromise on $x_R$.
We note that the qualitative behavior of the double-sinh $R$ shown is typical
for this CAP.

From the figure,
the similarity of the two reflection coefficients at low energies is evident.  Specifically, the
value of $K_\text{min}$\,---\,which has the biggest  influence on $x_R$\,---\,is
nearly identical between the two.  In fact, the 
optimal values of $\lambda_i$ from  the double-exponential CAP provide a very
good initial guess for the optimization of the double-sinh CAP.  

Also evident from Fig.~\ref{ds} is the difference in the two CAPs' $R$ at high energies.
Where the double-exponential $R$ rises back towards unity at high energies, the double-sinh
$R$ asymptotes to a value less than unity.  This value can be approximately calculated,
by considering only the $-i \alpha_2^2/x^2$ behavior of $V$, to be
\begin{equation}
R\xrightarrow[K\rightarrow\infty]{} e^{\pi \text{Im}\sqrt{1-4 i \lambda_2^2}},
\label{ds_HighEnergyR}
\end{equation}
consistent with the limiting behavior found in Ref.~\cite{Manolopoulos_JCP_2002}. 

Given the discussion in Sec.~\ref{ComplexPot}, one might wonder whether allowing the $\lambda_i$ 
to be complex\,---\,rather than real as assumed so far\,---\,could improve the CAPs'
performance even further.  The simple answer is that it can.  In fact, the
double-sinh CAP plotted in Fig.~\ref{ds} has complex $\lambda_i$.  We could
not, however, find a more optimal double-exponential CAP by allowing $\lambda_i$
to be complex for the present $E_\text{min}$, $E_\text{max}$, and $R_c$ (see, however,
Sec.~\ref{OtherRc}).

Incidentally, Eq.~(\ref{ds_HighEnergyR}) gives the reflection coefficient at {\em all}
energies for a CAP that has the form $-i \alpha_2^2/x^2$ everywhere (see App.~\ref{DipoleCAP}).  In particular,
the reflection coefficient is not unity at zero energy like the other CAPs we consider and thus
corresponds to $K_\text{min}$=0.  In many ways, such a CAP
would be ideal\,---\,no optimization would be necessary and $\lambda_2$ could simply be
calculated by setting Eq.~(\ref{ds_HighEnergyR}) to $R_c$.  Unfortunately, $x_R$=110\,a.u.
for such a CAP, rendering it uncompetitive with our best CAP\,---\,although better than  
the quadratic CAP  often used in the literature (see Table~\ref{optParams}).

One possible solution would be to simply cut the $-i \alpha_2^2/x^2$ CAP off at some $x_0$.
Intuitively, this should affect the low-energy behavior of $R$ for wavelengths comparable
to and larger than $x_0$.  The reflection coefficient in this case is again 
analytic (see App.~\ref{DipoleCAP}), and it can be seen after some exploration that while this
expectation is true, $R$ falls off at small $k$ more-or-less like $1/(kx_0)^4$ rather than like the exponential decrease of 
our best CAPs.  Since one chooses $x_0$ for this CAP from 
\begin{align}
 R(k_\text{min}x_0)\sim 1/(k_\text{min}x_0)^4=R_c,
\end{align}
$x_0$\,---\,and thus $x_R$\,---\,winds up being large.  For instance, $x_R$=57\,a.u.
for $R_c=10^{-3}$, which is about double that for our best CAP.

In the context of this discussion, the double-sinh CAP can be seen as providing
a smooth cutoff of the $-i \alpha_2^2/x^2$ CAP and similarly leads to modifications of Eq.~(\ref{ds_HighEnergyR})
at small energies.

\subsubsection*{Fall-to-the-center problem}

Whenever an attractive $1/x^2$ potential is used, one must take care to consider
the ``fall-to-the-center'' problem.  The real-valued version of such potentials are known~\cite{Landau} to 
have an infinite number of bound states with energies stretching to $-\infty$\,---\,a fact reflected in the wave 
function's oscillating an infinite number of times as $x\rightarrow 0$\,---\,so long
as the potential strength exceeds a critical value.  This is the quantum-mechanical
analog of the classical fall-to-the-center problem in such potentials.  Moreover,
this effect is possible even for potentials that are only $1/x^2$ for small $x$
like our double-sinh CAP.

No finite numerical representation\,---\,such as the grid methods common for
TDSE solvers\,---\,can represent the infinity of oscillations in
the fall-to-the-center regime, and any attempt to accurately represent even a finite number
of them will be very costly computationally.  

To understand how to avoid this regime, we must examine the small-$x$ behavior
of the wave function.  From App.~\ref{DipoleCAP} and using its notation, we see that
\begin{equation*}
\psi \xrightarrow[z\rightarrow 0]{} z^{\frac{1}{2}+(a_r-|a_i|)/\sqrt{2}} \exp\biggl[i\frac{a_r+|a_i|}{\sqrt{2}}\log z\biggr].
\end{equation*}
This solution assumes $a_r>|a_i|$ and shows that even for a nearly purely imaginary
CAP, $a_i$=0, the wave function oscillates an infinity of times as $z\rightarrow 0$.
Empirically, choosing $a_r\gg|a_i|$ so that the first term above suppresses the wave function
for $z\rightarrow 0$ is sufficient to prevent numerical difficulties.
Consequently, we have chosen $a_i$=0, which is equivalent to $\lambda_2^2=a_r^2-i/4$.

\subsection{Complex boundary condition}

We have so far assumed that the wave function vanishes on the boundary at $x=0$
as is typical for most TDSE solvers.
But, if the numerical method used to solve the TDSE is flexible enough to allow
complex log-derivative boundary conditions, then additional absorption can be built 
in at very little additional cost.

The effect of the complex boundary condition,
\begin{align} 
 \frac{1}{\psi} \frac{d\psi}{dx} = b,
\label{compBC}
\end{align} 
can be most easily illustrated for a free particle.  If one imposes Eq.~(\ref{compBC}) 
at $x$=0 as in Fig.~\ref{scheme}, but with no potential, one obtains the reflection
coefficient (see App.~\ref{BCAppend} for more details, including the effect on bound-state
energies)
\begin{equation}
R=\biggl|\frac{B+iK}{B-iK}\biggr|^2
\end{equation}
with $B=b\beta$.
To have absorption, we must have $\text{Im}\,B\leqslant 0$; to have maximum absorption,
we must have $\text{Re}\,B$=0.  Thus, setting $B=-iK_0$, we see that $R=0$ at $K=K_0$.
Such a boundary condition therefore makes the boundary perfectly transparent to an incident
plane wave of momentum $K_0$ and partially transparent to other momenta.  Moreover, it
does so with $x_R$=0.

Unfortunately, this boundary condition by itself cannot compete with the CAPs since $R$
cannot be made small enough over a large enough energy range.  Since implementing it, in principle,
requires no change in the spatial grid, though, the possibility of combining it with
a CAP and reducing $x_R$ further is worth exploring.

At low energies, the CAP will dominate the behavior of $R$, and the boundary condition will
have little influence.  Therefore, one should try to use the boundary condition to  modify
the high-energy $R$ where it can dominate the behavior.  In general, choosing $B\sim -iK_\text{max}$
is a good initial guess and allows the reduction of $\lambda$\,---\,and therefore $x_R$.

It should be noted that a complex boundary condition cannot be used with the
double-sinh CAP due to  its singularity at the boundary.  Like the centrifugal
potential that it resembles, the double-sinh CAP has one regular solution that vanishes
at the boundary and one irregular solution that diverges at the boundary~\cite{Manolopoulos_JCP_2002}.
Therefore, it is not possible to form the necessary linear combinations to satisfy
Eq.~(\ref{compBC}).

\subsubsection{Single-exponential CAP}

The reflection coefficient for a single-exponential CAP with a complex boundary condition
is again analytic and is given in Eq.~(\ref{seR_BC}).  The CAP parameters must be re-optimized
along with the value of $b$, and the procedure is largely the same as described above.  The
fact that $K_\text{min}$ is essentially unaffected by the addition of the complex boundary
condition\,---\,so long as $|B|\sim|K_\text{max}|$\,---\,simplifies the process.

\begin{figure} 
\includegraphics[width=\columnwidth]{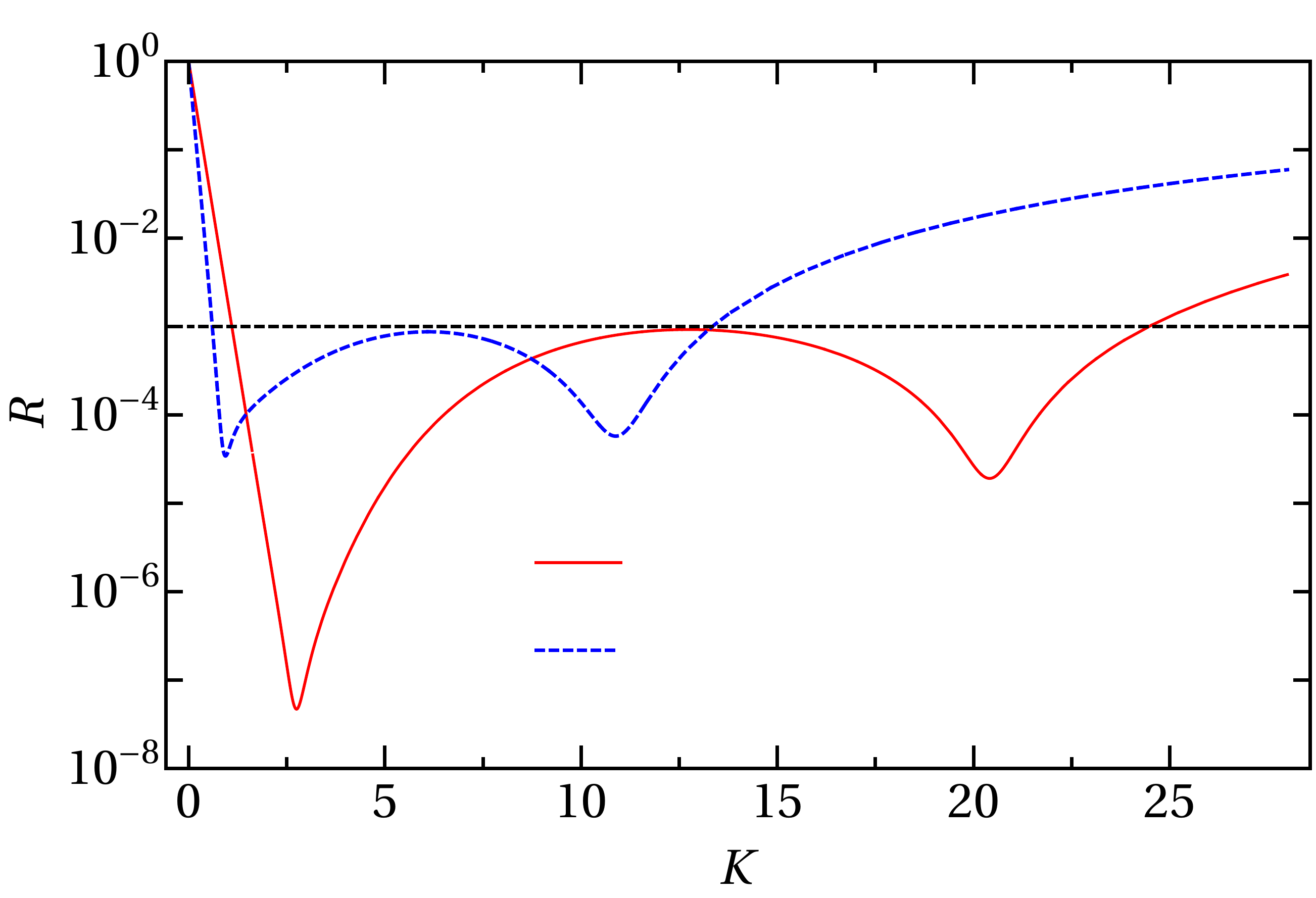}
\setlength{\unitlength}{0.01\columnwidth}
\begin{picture}(0,0)(50,0)
\put(49,29){\scalebox{1.0}{$\lambda^2=26i,B=-20.4i$}}
\put(49,22){\scalebox{1.0}{$\lambda^2=40.1e^{i0.11\pi}, B=-12.5i$}} 
\put(20,50){\scalebox{1.0}{$R_c$}}  
\end{picture}
\caption{Reflection coefficient as a function of the unitless momentum for a
single-exponential CAP with complex boundary conditions: purely imaginary $\lambda^2$
and complex $\lambda^2$.} 
\label{seOpt_BC}
\end{figure}
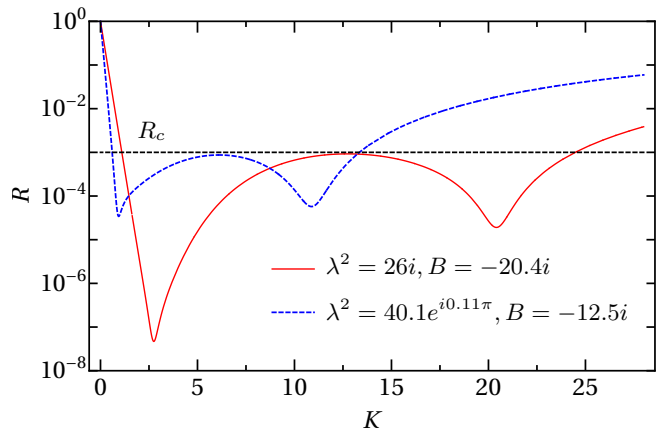

Examples of optimal choices are shown in Fig.~\ref{seOpt_BC} for a purely imaginary CAP
and a complex CAP.  Comparison with the reflection coefficients shown in Sec.~\ref{seCAP}
shows the effect of the complex boundary condition through the appearance of the high-energy minimum
close to $K=|B|$.  In both cases, the complex boundary condition has produced a roughly 15\% 
reduction in $x_R$ to 71.7\,a.u. and 48.1\,a.u., respectively.

\subsubsection{Double-exponential CAP}

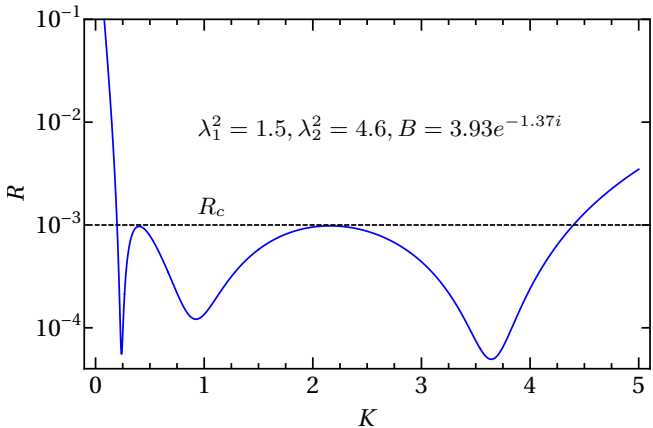
\begin{figure} 
\includegraphics[width=\columnwidth]{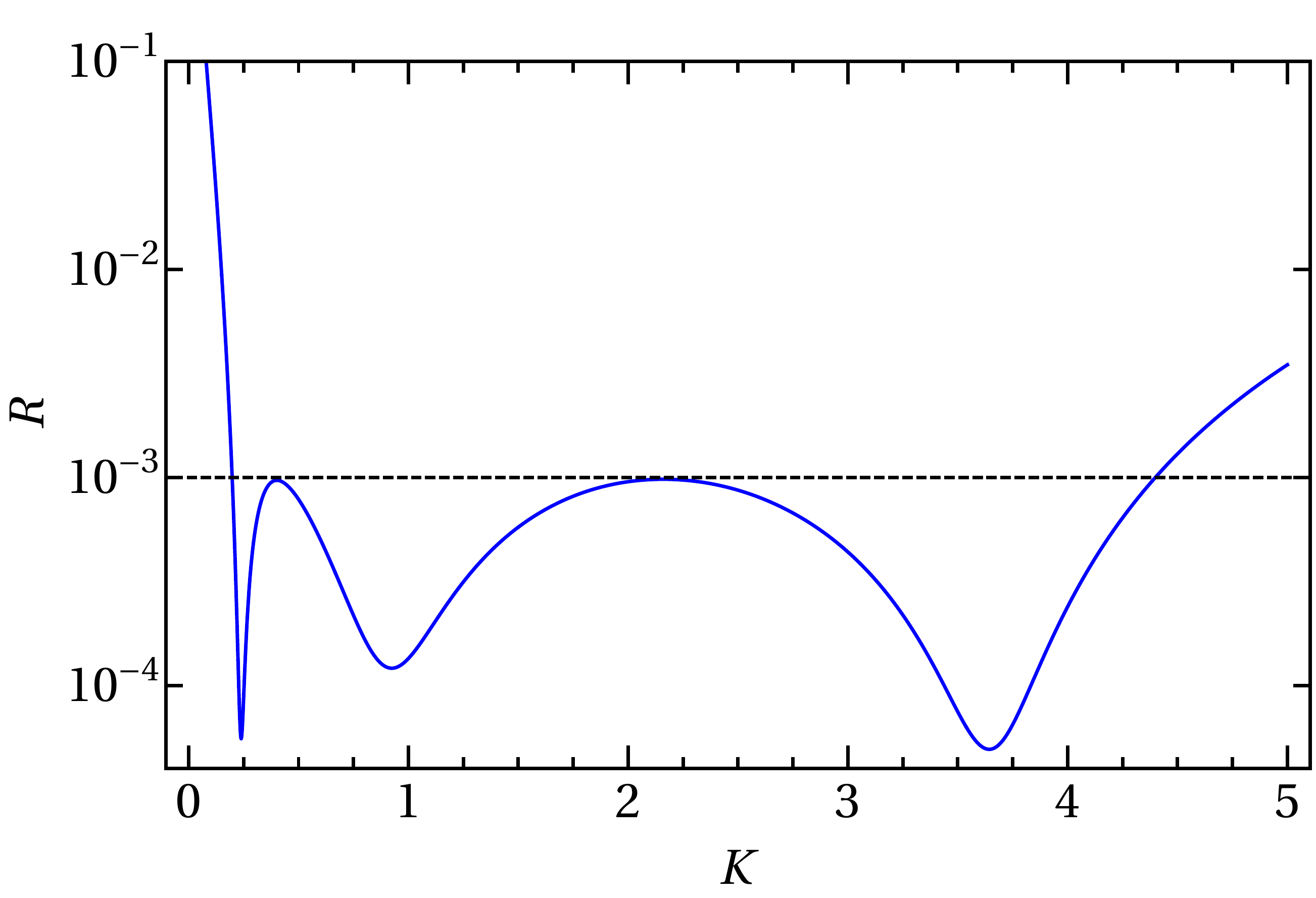}
\setlength{\unitlength}{0.01\columnwidth}
\begin{picture}(0,0)(50,0) 
\put(30,50){\scalebox{1.0}{$\lambda_1^2=1.5,\lambda_2^2=4.6, B=3.93e^{-1.37i}$}} 
\put(30,38){\scalebox{1.0}{$R_c$}}
\end{picture}
\caption{Optimum reflection coefficient as a function of the unitless
momentum for a double-exponential CAP with complex boundary
conditions.
} 
\label{deOpt_BC}
\end{figure}

Adding a complex boundary condition to the double-exponential CAP also produces an
analytic expression for $R$ as given in Eq.~(\ref{deR_BC}).  Re-optimizing the 
parameters yields the reflection coefficient shown  in Fig.~\ref{deOpt_BC}.  As
with the single-exponential CAPs, the boundary condition has introduced a high-energy
minimum near $K=|\text{Im}\,B|$.  Unlike the single-exponential CAPs, though, the
minimum $x_R$ was found for $\text{Re}\,B\neq 0$.

This optimum double-exponential CAP continues the pattern that has emerged as we
have found improved CAPs: namely, that we add more structure to $R$ and decrease
the absorption for the mid-range of $K$.  The double-exponential-CAP reflection
coefficients in Fig.~\ref{de_Opt}, for instance, have comparatively little structure\,---\,mainly
a minimum in $R$.  Moreover, this minimum is relatively broad and orders of magnitude
lower than $R_c$.  The $R$ shown  in Fig.~\ref{deOpt_BC}, in contrast, has three
narrower minima only one order of magnitude or so lower than $R_c$.

\section{Optimal CAP} 
\label{CAPComp}

To determine which CAP\,---\,among those listed in Table~\ref{CAPs}\,---\,is the best,
we numerically searched for their optimal parameters, assuming they are purely imaginary potentials.
From  the discussion above, we know that each could be improved by including a real
potential and a complex boundary condition, but we expect\,---\,and confirmed with
spot tests\,---\,that the relative performance of the CAPs will remain 
qualitatively the same.
As mentioned previously, we selected the CAPs to compare based on their apparent
popularity in the literature or on the claims made for their performance.

In optimizing these CAPs, we follow the principles described in previous sections that 
the width of the CAP determines the long-wavelength absorption; and the depth, the 
short-wavelength. The optimization is then reasonably straightforward once we identify 
the parameters corresponding to the width and depth.

\begin{table*}
\caption{Comparison of the optimal absorption ranges for all the CAPs considered.
The optimal parameters are given for the electron in our strong-field application\,---\,see
Eqs.~(\ref{Params}) and (\ref{MomRange})\,---\,so that all quantities are in atomic units. 
} 
\begin{ruledtabular} 
\begin{tabular}{l c  c cc r} 
 CAP type \rule{0pt}{2.4ex}  & $\alpha^2$ or 
                    ($\alpha_1^2$, $\alpha_2^2$)    	& $b$           & $\beta$ & $x_0$   & $x_R$ \\  \hline
 quadratic \rule{0pt}{2.6ex} & $1.21\times 10^{-5}$&               & ---     & 129     & 124  \\ 
 cosine masking function     & 15.9                &               & 810     & 128     & 124\\ 
 M-JWKB                      & ---                  	&               & ---     & 119     & 118 \\
 quartic                     & $2.40\times10^{-9}$ &               & ---     & 112     & 95 \\ 
 pseudo-exponential          & $4.54\times10^5$   	&               & $3.27\times10^{3}$   & 240     & 88 \\
 P\"oschl-Teller             & 11.0                 &               & 20.3    & 40.0    & 85 \\ 
 single-exponential          & $0.849i$             &               & 10.0    & ---     & 84 \\ 
 single-exponential+BC       & $0.260i$             & $-2.04i$      & 10.0    & ---     & 72 \\ 
 single-exponential          & $5.24e^{i0.11\pi}$   &               & 5.62    & ---     & 57 \\ 
 single-exponential+BC       & $1.35e^{i0.11\pi}$   & $-2.29i$      & 5.45    & ---     & 48 \\ 
 double-exponential          & (0.839,5.09)         &               & 1.79    & ---     & 30        \\
 double-sinh                 & ($0.298e^{0.104i}$,$0.71e^{-0.0906i}$)   && 1.97    & ---     & 29 \\
 double-exponential+BC       & (0.463,1.42)         & $2.19e^{-1.37i}$  & 1.80    & ---     & 28 
\end{tabular} 
\end{ruledtabular} 
\label{optParams} 
\end{table*} 

\begin{figure} 
\includegraphics[width=\columnwidth]{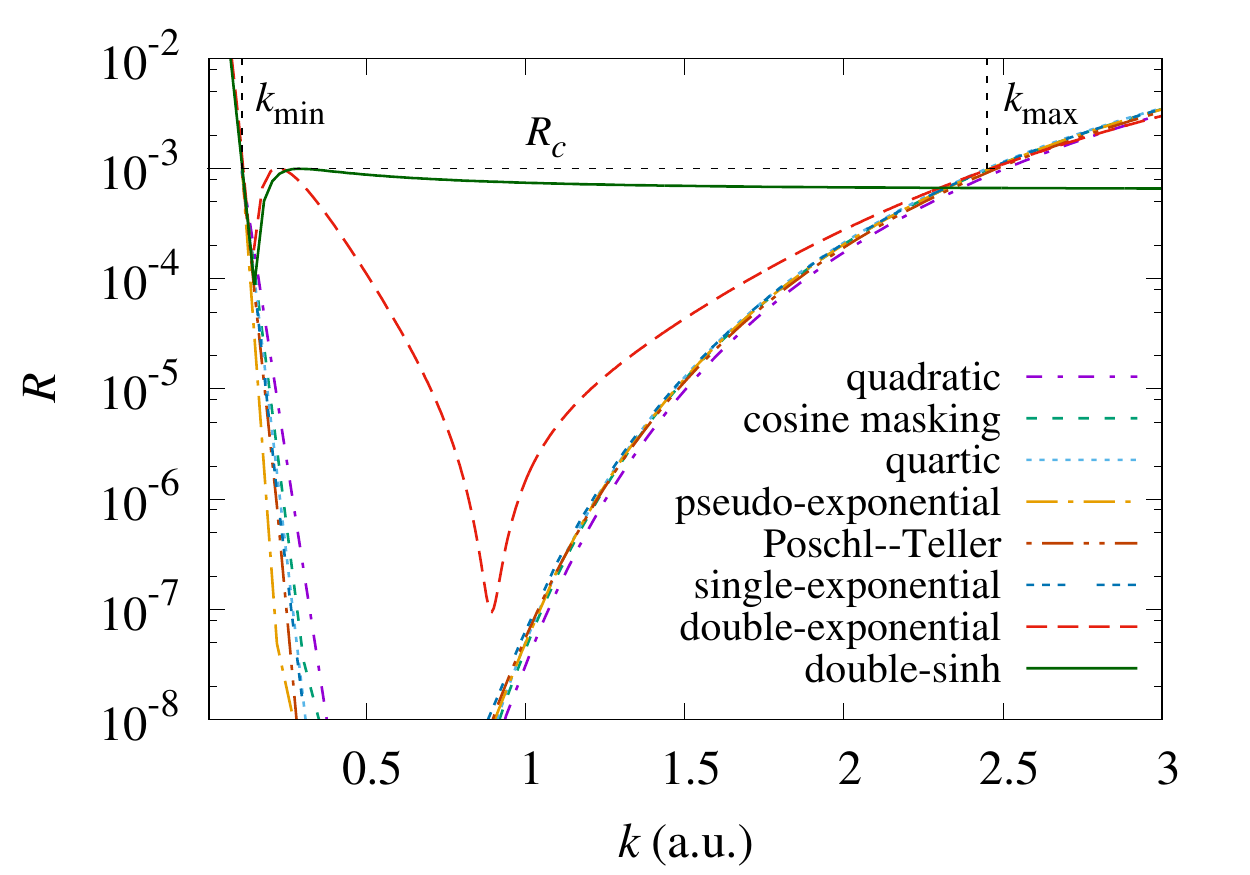}
\caption{\label{RC} 
Optimal reflection coefficients for all CAPs as a function of the electron's momentum
using the parameters from Table~\ref{optParams}. They all satisfy the criteria that 
$R \leqslant R_c$ for $0.006 \leqslant E \leqslant 3~{\rm a.u.}$, as required. } 
\end{figure}

In Table~\ref{optParams}, we list the optimal parameters we have found for our $E_\text{min}$,
$E_\text{max}$, and $R_c$.  The table includes the resulting values of $x_R$, and we expect 
that they are the optimal values to within a few percent. Note that we used $\delta$=0.1 for
the M-JWKB CAP based on the solution of $R(E_\text{min})=R_c$ taken from Fig.~3 of Ref.~\cite{Manolopoulos_JCP_2002}.
We show in Fig.~\ref{RC} the corresponding reflection coefficients.

The cosine masking function should be regarded as a polynomial CAP 
since its behavior in $0\leqslant x\leqslant x_0$ for the optimal parameters of 
Table~\ref{optParams} is largely indistinguishable from the quadratic CAP\,---\,thus
its $x_R$ is identical to the quadratic CAP.  Similarly, for the optimal
parameters we found, only the exponential tail of the P\"oschl-Teller potential
remained on the grid, making its performance essentially identical to  that
of the purely imaginary single-exponential CAP.

The absorption ranges $x_R$ listed in Table~\ref{optParams} display a 
surprisingly large range\,---\,more than a factor of 4. 
Comparing only the purely imaginary potentials, the exponential and P\"oschl-Teller forms are more 
than 30\% more efficient than the quadratic CAP.  They are also more efficient than
the quartic CAP.  So, while the exponential form generally seems better, the majority
of the disparity in Table~\ref{optParams} arises from adding a real part to the  CAP
and imposing a complex boundary condition.  

From Table~\ref{optParams}, the best performance is found for the double-exponential and double-sinh CAPs, outperforming 
the next-best CAP\,---\,the complex-valued, complex-boundary-condition, single-exponential CAP\,---\,by roughly
40\%.  Compared to the next-best purely imaginary CAP, they hold nearly a factor of 3 advantage in $x_R$.
For reference, we tested the strategy of adding a real part and a complex boundary condition to the 
quadratic CAP and found $x_R$ shrank only to about 70\,a.u.  So, while pursuing this strategy with
the other CAPs in the table would reduce their $x_R$,
we believe the double-exponential and double-sinh CAPs would remain the best. 
Interestingly, since the de~Broglie wavelength at $k_\text{min}$ is 57\,a.u., our best CAP
manages its efficient absorption in a range of only about half this longest wavelength.

Our recommendation, therefore, is to use the double-sinh CAP when its singularity at the
boundary causes no numerical difficulties.  In the cases that it does, then the double-exponential
CAP is the best choice.  The remainder of our discussion will  thus focus on  these two
CAPs.

\section{Other absorption criteria}
\label{Other}

The discussion and optimization so far has centered on the $E_\text{min}$,  $E_\text{max}$, and $R_c$
from Eq.~(\ref{Params}).  Other choices may well be needed, however, for other choices
of laser parameters or calculational goals.  We thus present
in this section the optimal parameters for the double-exponential and double-sinh CAPs for a
selection of likely changes in $E_\text{min}$,  $E_\text{max}$, and $R_c$.

\subsection{Different energy range}

\subsubsection{Changing $E_\text{max}$}

Computationally, the main challenges in solving the 
TDSE\,---\,especially for current and future laser parameters of experimental 
interest\,---\,are that in the course of its strongly-driven dynamics, the electron travels far from the 
nucleus and gains substantial energy.  In particular, we still expect $E_\text{max}\propto U_p \propto I/\omega^2$,
so that it will grow either with increasing intensity or increasing wavelength\,---\,both of which
are certainly of interest.  While $E_\text{min}$ does not change in this case,
$E_\text{max}$ does, and the CAP must accommodate it.  

Under these conditions, the double-sinh CAP from Table~\ref{optParams} works without change since it
has no $E_\text{max}$.  In fact, this is its primary advantage.  The double-exponential CAP,
on the other hand, must be re-optimized for each $E_\text{max}$.  As discussed in Sec.~\ref{Method},
$\lambda_2$ needs the greatest changes\,---\,but should have minimal impact on $x_R$\,---\,and 
these expectations are reflected in the optimal parameters
shown in Table~\ref{longwavelength} for two longer wavelengths.  These parameters
were found following the same procedure as above with the same $k_\text{min}$ and $R_c$
and with $E_\text{max}$=$10\,U_p$ at $10^{14}$\,W/cm$^2$.  They were found assuming $\psi=0$
on the boundary, but parameters could certainly be found for a complex boundary condition as well.
Note that $x_R$ changes less than about 10\% as expected.

\begin{table} 
\caption{\label{longwavelength} Optimal parameters for the double-exponential CAP for an
electron exposed to longer wavelengths.
Per the discussion  in the text, the only impact of wavelength here is on $E_\text{max}$.
All quantities are  in atomic units unless otherwise specified.} 
\begin{ruledtabular}
\begin{tabular}{cc c cccc} 
 $\lambda$ (nm) & $E_{\rm min}$   & $E_{\rm max}$   & $ \alpha_1^2$ & $\alpha_2^2$  & $\beta$ & $x_R$ \\ \hline 
 800 \rule{-3pt}{2.6ex}  & 0.006 & 3   & 0.839 & 5.09 & 1.79 & 29.9  \\ 
 1600   & 0.006 & 8.8   & 1.07 & 8.37 & 1.79 & 30.7  \\ 
 2400   & 0.006 & 20 & 1.31 & 12.4 & 1.79 & 31.5 
\end{tabular} 
\end{ruledtabular} 
\end{table} 
 
\subsubsection{Changing $E_\text{min}$}

In our optimization scheme above, we set $E_\text{min}$ to 0.1\,$\hbar\omega$ for 800-nm light.
This choice was motivated by the need to ensure that the entire ionized electron wavepacket
is absorbed by the CAP.  However, the CAP is often placed at a large distance from the 
nucleus so that these very slow electrons may not have time to reach the CAP during the
propagation.  In this case, $E_\text{min}$ can be increased, thereby decreasing $x_R$.

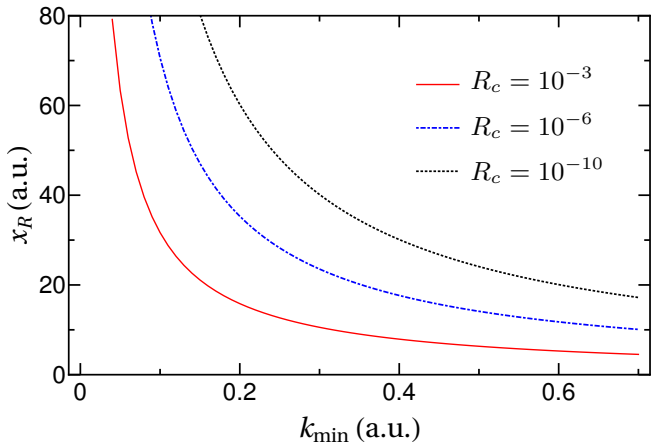
\begin{figure}[h] 
\includegraphics[width=\columnwidth]{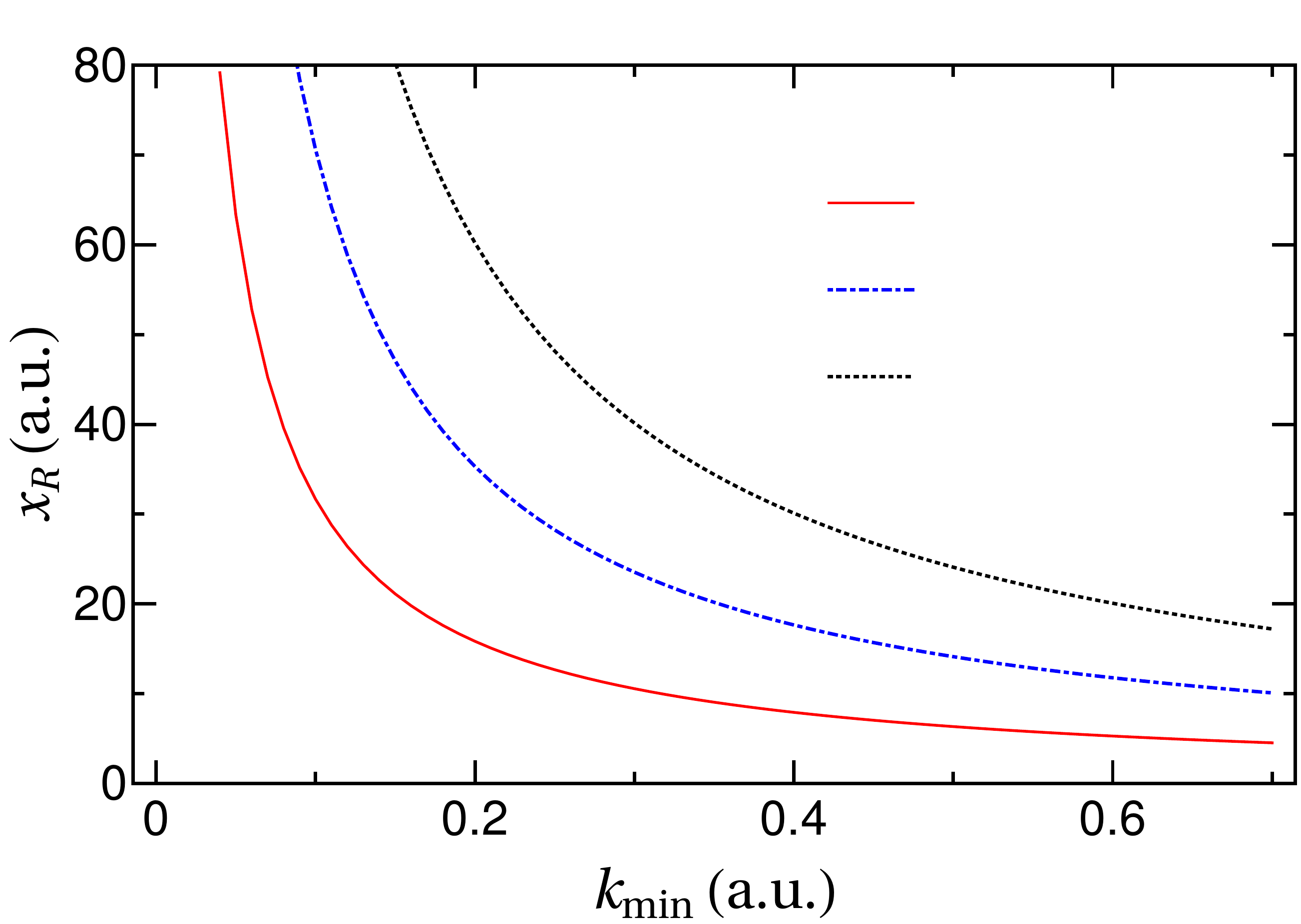}
\setlength{\unitlength}{0.01\columnwidth}
\begin{picture}(0,0)(50,0) 
\put(72,59){\scalebox{1.1}{$R_c=10^{-3}$}}  
\put(72,52){\scalebox{1.1}{$R_c=10^{-6}$}}
\put(72,45){\scalebox{1.1}{$R_c=10^{-10}$}}
\end{picture}
\caption{Absorption range $x_R$ as a function of $k_\text{min}$ for the double-sinh CAP.
The parameters for $R_c \leqslant 10^{-3}$ can be found in Table~\ref{OtherRcTable}.
}
\label{ds_xR}
\end{figure}

Modifications to $E_\text{min}$ for the double-sinh CAP are straightforward and do not
require re-optimization\,---\,again, thanks to the lack of an $E_\text{max}$.  Changing
$k_\text{min}$ just means recalculating $\beta$ using $\beta=K_\text{min}/k_\text{min}$
since $K_\text{min}$ is fixed. Figure~\ref{ds_xR} shows the $x_R$ that results as a function
of $k_\text{min}$.  The figure shows that for modest increases in $k_\text{min}$ from our
choice in Eq.~(\ref{MomRange}), $x_R$ can be decreased substantially.  For example, for $k_\text{min}$ 
above about 0.3\,a.u., $x_R$ is smaller than 10\,a.u. for $R_c$=$10^{-3}$. For $k_\text{min}$
above about 0.4\,a.u., the $x_R$ for $R_c$=$10^{-10}$ is equal to or smaller than the 
original $x_R$=28.8\,a.u. for the double-sinh CAP.

For a double-exponential CAP, it is still true that
the larger $k_\text{min}$, the smaller $\lambda_1$ and $\lambda_2$, and the smaller $x_R$. 
However, re-optimization is required to obtain
the smallest $x_R$.  For instance, if one can accept doubling $k_{\rm min}$ to 0.22\,a.u.,
then we can have 
\begin{equation} 
 \lambda_1^2 = 2.00  \text{ and } \lambda_2^2 = 9.11,   \text{ so that }x_R =17.0\,\text{a.u.}
\label{Higherkmin}
\end{equation} 
with $\beta$=0.90\,a.u.  

On the other hand, the double-exponential CAP can be adjusted just like the 
double-sinh CAP if a less-than-optimal $x_R$ can be tolerated.  Specifically, 
the values of $\lambda_i^2$ can be kept, so that $K_\text{min}$ does not change, and $\beta$
can be recalculated from $\beta=K_\text{min}/k_\text{min}$.  In this case, $k_\text{max}$
grows to $k_\text{min} K_\text{max}/K_\text{min}$, guaranteeing absorption at the
prescribed level beyond $E_\text{max}$.  The resulting $x_R$ looks very much like
those in Fig.~\ref{ds_xR}, except that $x_R$ for $R_c$=$10^{-3}$ does not go below
10\,a.u. until $k_\text{min}$=0.45\,a.u.  For comparison, $x_R$=17.4\,a.u. at $k_\text{min}$=0.22\,a.u.
and is thus slightly worse than the fully re-optimized result in Eq.~(\ref{Higherkmin}).

\subsection{Different $R_c$}
\label{OtherRc}

One of the primary design goals of a CAP is to leave the physical wave function\,---\,the wave
function outside the absorption region\,---\,unaffected.  Of course, this goal can only be achieved 
to a given accuracy, and that accuracy is controlled by $R_c$.  To see the relation, consider the physical wave 
function written in Fig.~\ref{scheme} from which $R$ is extracted,
\begin{equation}
\psi(x) = e^{-ikx}+\sqrt{R}e^{i\varphi}e^{ikx}, ~ x\geqslant x_R.
\label{TIwfn}
\end{equation}
The second term is precisely the unwanted contribution from reflection, and it is limited 
by $R\leqslant R_c$ by design.  Given that this is just one component of the time-dependent wave
function, this error is always relative to  the first term.  In other words, if one desires
$n$ digits to be accurate, then one should choose $R_c$=$10^{-2n}$.

We thus provide in Table~\ref{OtherRcTable} the optimal
parameters for the double-exponential and double-sinh CAPs with $\psi$=0 on the boundary, assuming $E_\text{min}$=0.006\,a.u.
and $E_\text{max}$=3\,a.u. as before, for several smaller $R_c$.  These results show that the absorption range for each type of CAP
is comparable, with the double-sinh CAP tending to be a few percent smaller.
Qualitatively, the reflection coefficients as a function of $K$ resemble those shown previously for $R_c$=$10^{-3}$.
As with the other CAP parameters we have given, we expect these to produce $x_R$ within a few percent of
the global optimum.

\begin{table*}
\caption{\label{smallerRc} Optimal parameters for double-exponential and double-sinh CAPs
for $R_c \leqslant 10^{-3}$.}
\begin{ruledtabular}
\begin{tabular}{ccccccccc}
\multicolumn{4}{c}{Double exponential} &&\multicolumn{4}{c}{Double sinh} \\ \cline{1-4}\cline{6-9}
 $\lambda_1^2$ & $\lambda_2^2$ & $\beta$\,(a.u.) & $x_R$\,(a.u.) & $R_c$\rule{-3pt}{2.8ex} & $x_R$\,(a.u.) & $\lambda_1^2$ & $\lambda_2^2$ & $\beta$\,(a.u.) \\ \hline
2.69 \rule{-3pt}{2.8ex}& 16.3 & 1.79 & 29.9  & $10^{-3}$ &
     28.8 & $1.16e^{0.104i}$ & $2.75-0.25i$ & 1.97 \\
$4.83e^{0.187i}$ & $31e^{0.0968i}$ & 2.77 & 44.7 & $10^{-4}$ &
     40.4 & $1.78 e^{0.23i}$ & $4.30-0.25i$ & 2.90  \\
$7.21e^{0.0486i}$ & $57.6e^{-0.411i}$ & 3.38 & 54.5 & $10^{-5}$ &
     52.6 & $2.66e^{0.346i}$ & $6.82-0.25i$ & 3.87  \\
$16.1e^{-0.219i}$ & 80.0 & 4.02 & 68.4 & $10^{-6}$ &
     64.2 & $3.8 e^{0.36i}$ & $9.67-0.25i$ & 4.77 \\
$19.4 e^{i0.135\pi}$ & $141e^{-0.073i}$ & 6.64 & 102 & $10^{-8}$ &
     89.4 & $6.75 e^{0.46i}$ & $17.2-0.25i$ & 6.77 \\
$30.8e^{i0.132\pi}$ & $232e^{-0.204i}$ & 8.48 & 130 & $10^{-10}$ &
     109 & $11.85 e^{0.14i}$ & $26.9-0.25i$ & 8.02 \\
48 & $355e^{-0.328i}$ & 9.04 & 144 & $10^{-12}$ &
     132 & $16.1 e^{0.21i}$ & $38.7-0.25i$ & 9.77
\end{tabular}
\end{ruledtabular}
\label{OtherRcTable}
\end{table*}

Note that the probability density corresponding to the wave function in 
Eq.~(\ref{TIwfn}),
\begin{equation}
|\psi(x)|^2 = 1+R+2\sqrt{R} \cos(2kx+\varphi),
\label{probdensity}
\end{equation}
can be useful for diagnosing issues with the CAP in a time-dependent calculation.  In particular, the last 
term above is the source of the telltale ripples in the probability 
density near the edge of a grid.  The ripples' size is limited by $\sqrt{R_c}$
and identifying their wavelength via Eq.~(\ref{probdensity}) in a time-dependent wave function reveals
the offending energy.

\section{\label{TDSE} Time-Dependent Demonstration} 

\begin{figure} [h] 
\includegraphics[width=\columnwidth]{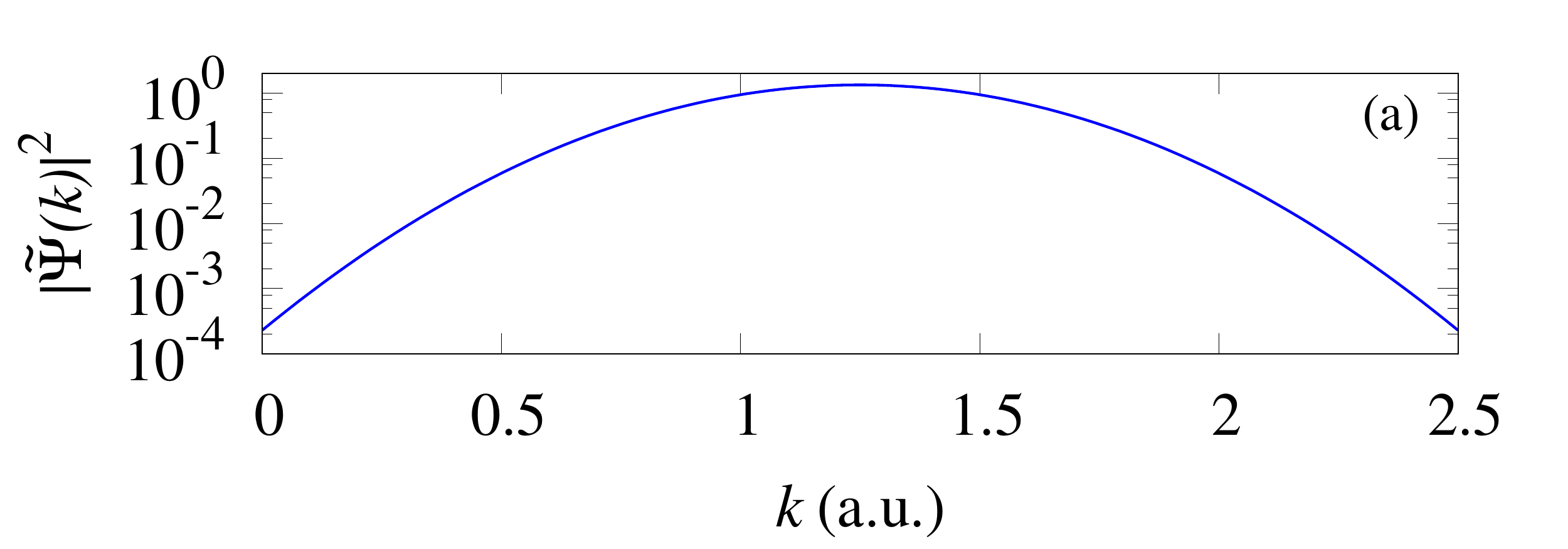}
\includegraphics[width=\columnwidth]{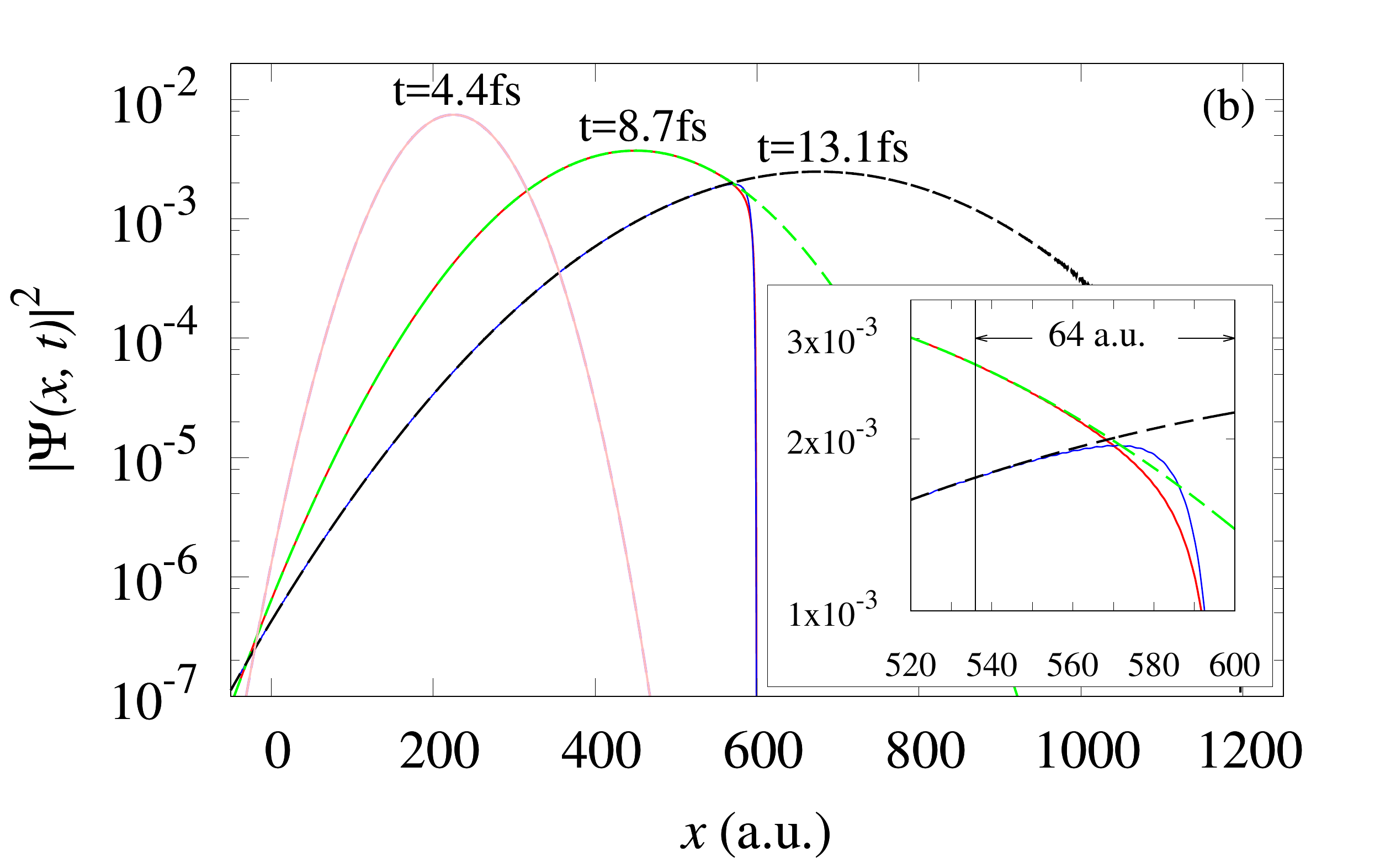}
\caption{\label{TDDemo} 
(a) Momentum distribution of the free wavepacket, covering $0.11$$\leqslant$$k$$\leqslant$$2.45\,\text{a.u.}$
(b) Demonstration of the $R_c$=$10^{-6}$ double-sinh CAP using a free wavepacket.
Solid lines show the with-CAP wavepacket, calculated for $-600\,\text{a.u.}$$\leqslant$$x$$\leqslant$$600\,\text{a.u.}$;
and dashed lines, the without-CAP wavepacket, calculated for $-600\,\text{a.u.}$$\leqslant$$x$$\leqslant$$1200\,\text{a.u.}$
Inset: Expanded view of the absorption region $536\,\text{a.u.}$$\lesssim$$x$$\leqslant$$600\,\text{a.u.}$ 
for clearer comparison. 
} 
\end{figure}

To verify that the improved performance of our recommended CAPs does indeed carry over to the
time-dependent problem and its numerical solution, we solve the TDSE for free-electron wavepacket propagation.
The wavepacket we use possesses a broad momentum distribution comparable to the 
target momentum range from Eq.~(\ref{MomRange}), as shown in Fig.~\ref{TDDemo}(a).

We again use FEDVR as the spatial representation and propagate the wave function
using the short-time evolution operator
\begin{equation}
\psi(x,t+\delta) = e^{-iH\delta} \psi(x,t)
\end{equation}
where the Hamiltonian includes the CAP $V(x)$,  
\begin{equation}
H=H_0+V,
\end{equation}
and $H_0$ is merely the kinetic energy in the present case.
We evaluate $e^{-iH\delta}$ using the split-operator form~\cite{Schneider_QDI_2011}
\begin{equation}
e^{-iH\delta} \approx e^{-iV\frac{\delta}{2}}e^{-iH_0\delta}e^{-iV\frac{\delta}{2}}.
\label{SplitOp}
\end{equation}
Since $V$ is diagonal in FEDVR, $e^{-iV\delta/2}$ can be easily evaluated and applied.
Moreover, in this form, the singularity in  the double-sinh CAP causes no problems
whatsoever.  The action of the remaining term in $H_0$ is calculated via a Pad\'e approximation~\cite{BLANES_PhysRep_2009}.

Equation~(\ref{SplitOp}) is a simple and convenient way to add a CAP to any propagator.
In many cases, the alternative, keeping the CAP in $H$, would require modifications of the propagation
algorithm or parameters to handle its non-Hermiticity or the singularity of the double-sinh CAP\,---\,or both.
These issues were discussed somewhat in Sec.~\ref{dsCAP} and more in Ref.~\cite{Manolopoulos_JCP_2002}.
Using Eq.~(\ref{SplitOp}) avoids these concerns and is more than sufficient for the application of the CAP.

The FEDVR element distribution is uniform in the range $-600\,{\rm a.u.} \leqslant x  \leqslant 1200\,{\rm a.u.}$,
and we require $\psi=0$ at the boundaries. 
Given that the wavepacket is initially centered near $x=0$, this box is 
large enough for the wavepacket to propagate for 10\,fs without significant reflection at the
boundaries even without CAPs. This will be our reference solution.
We carry out a second, identical propagation but apply the double-sinh CAP at 
$536\,\text{a.u.}   \lesssim x  \leqslant  600\,\text{a.u.}$.  For this example, we choose the 
CAP designed to have $R_c=10^{-6}$ using the parameters shown in Table~\ref{smallerRc}.
We thus compare the wavepacket with and without applying the CAP. 
All the results have been tested to be converged to at least 3 digits with respect to
all numerical parameters.

Figure~\ref{TDDemo} shows the two wavepackets at different times.  It is clear 
that the CAP is performing as expected since the wavepacket decays in the absorbing region
without any of the characteristic oscillations of reflections visible\,---\,at least without enlarging the
plot by a factor of four or five.  For comparison, the wavepacket without
a CAP equally clearly shows the reflection oscillations at $t$=13.1\,fs for reflections from
the boundary at $x$=1200\,a.u. 

In addition, the with-CAP wavepacket is
numerically unaffected before entering the absorbing region, agreeing with the without-CAP
wavepacket to at least 3 digits for $x \lesssim 536$\,a.u. for all times, even as more than 70\%
of the wavepacket has been absorbed. This agreement shows that the absorption range $x_R$
defined in the time-independent study is consistent with the results from the time-dependent 
propagation.  

Enlarging Fig.~\ref{TDDemo} by a factor of at least four or five will reveal the tiny reflection ripples
in the with-CAP wavepacket near and in the absorption region.  Their relative magnitude is about $10^{-3}$=$\sqrt{R_c}$
as expected. Per the discussion in Sec.~\ref{OtherRc}, such oscillations are unavoidable with a CAP
and testing with other CAPs and values of $R_c$ further support the conclusions there. For example, the oscillation for 
$R_c$=$10^{-3}$ becomes fairly noticeable, 
which suggests that $R_c$ should in practice be no larger than $10^{-4}$\,---\,{\em i.e.}, two digits in
the wave function\,---\,to provide quantitatively reliable results.
Finally, we note that the roughly 15-fs propagation time is comparable to a typical strong-field 
calculation, bringing some realism to this simple demonstration.

\section{Summary} 
We have presented a systematic study to boost the performance of complex absorbing 
potentials. 
Based on the time-independent reflection coefficient, we were able 
to quantitatively design the most optimal absorbing potential of a given form.
In particular, for ultrafast, strong-field TDSE solvers, the optimal CAP parameters should be determined by 
the absorbing energy range required for the laser parameters and by the desired accuracy of 
the TDSE solutions.

We proposed two new CAPs\,---\,namely, the double-sinh CAP and the double-exponential 
CAP\,---\,that significantly outperform the CAPs currently in standard use.
Their superiority was demonstrated through comparison with optimized versions of most of the CAPs
commonly found in the literature.
Both of our proposed CAPs overcome the primary impediment to efficient performance\,---\,absorption of 
long wavelengths\,---\,while also absorbing a large range of energies that covers basically all 
strong-field processes.
Because we quantified the CAP's performance and identified $x_R$ as the figure of merit
for their efficiency, we could show that using an exponential CAP already improved on the common 
quadratic CAP's performance by one third. A further factor of almost three was gained, however, by
adding a real part to the CAP\,---\,a strategy well known in other uses of CAPs, but not in strong-field
applications. 

We highly recommend the double-sinh CAP for local spatial 
representations, such as FEDVR where the potential is diagonal.
It is efficient, easy to use, and easy to 
adapt to different absorption criteria\,---\,i.e., energy range and level of absorption.
Incorporating it into the time propagation via split-operator methods is easy and effective.

For other spatial representations, the double-sinh and the double-exponential CAPs are equally 
recommended. However, care might need to be taken for the double-sinh singularity close to the boundary. 
In case the double-sinh singularity is a problem for the propagator, 
the double-exponential CAP should be chosen. 
Although optimization of the double-exponential CAP is more involved than for the double-sinh
CAP, it is still fairly straightforward. Its optimization procedure, 
along with that for the double-sinh CAP, is detailed in this work.

\begin{appendices} 

\section{\label{expAppend}Reflection Coefficient for Single-Exponential CAP} 

We start from the Schr\"odinger equation for the single-exponential CAP: 
\begin{align} 
   \left[-\frac{\hbar^2}{2 m} \frac{d^2}{d x^2} - \frac{\alpha^2\hbar^2}{2 m} e^{-x/{\beta}} \right] \psi 
   = E \psi = \frac{\hbar^2  k ^2}{2 m} \psi.
\end{align} 
Setting $z= x/{\beta}$ gives
\begin{align} 
 \left[ -\frac{d^2}{d z^2} - \lambda^2 e^{-z} \right] \psi = K^2 \psi, 
 \label{expTISE} 
\end{align} 
where $\lambda \equiv \alpha \beta $ and $K \equiv k \beta$. 
We define 
\begin{align} 
   \xi = 2 \lambda e^{-z/2}, \nonumber 
\end{align} 
so that Eq.~(\ref{expTISE}) becomes  
\begin{align} 
 \left[ \xi^2 \frac{d^2}{d \xi^2}+ \xi \frac{d}{d \xi} + \xi^2 + 4 K^2 \right] \psi =0.  
\end{align} 
This is Bessel's equation; the general solution is thus
\begin{equation} 
 \psi = C J_{2 i K}(\xi) +D J_{-2 i K}(\xi). 
\label{se_wfn}
\end{equation} 

To obtain the reflection coefficient, we need $C$ and $D$ and we need to analyze the asymptotic behavior of these solutions. 
Starting with the latter,
the $z  \rightarrow   \infty $ ($x\rightarrow\infty$) behavior can be found from the $\xi \rightarrow 0$ expansions,
\begin{align} 
 J_{2 i K}(\xi)  &\xrightarrow[z \rightarrow \infty]   ~\frac{\lambda^{2 i K}}{\Gamma(1+2 i K)} e^{-i K z}  \nonumber \\ 
 J_{-2 i K}(\xi) &\xrightarrow[z \rightarrow \infty] ~\frac{\lambda^{-2 i K}}{\Gamma(1-2 i K)} e^{i K z}.
\end{align} 
To find $C$ and $D$, we impose the $x=0$ boundary condition,
\begin{align} 
 \psi(x=0) &= \psi(z=0) = \psi(\xi = 2 \lambda) = 0 .
\end{align} 
Thus, 
\begin{align} 
   D = -\frac{J_{2 i K}(2 \lambda)}{J_{-2 i K}(2 \lambda)} C. 
\end{align} 

Finally, the asymptotic solution reads 
\begin{multline} 
 \psi \xrightarrow[z \rightarrow \infty]~ C \biggl[ \frac{\lambda^{2 i K}}{ \Gamma(1+2 i K)} e^{-i K z }  \\
 -\frac{\lambda^{-2 i K}}{ \Gamma(1-2 i K)} \frac{J_{2 i K }(2 \lambda)}{J_{-2 i K }(2 \lambda)} e^{i K z } \biggr]. 
\end{multline} 
From this expression, the reflection coefficient can be found to be
\begin{align} 
  R= e^{4K \arg \lambda^2} \left| \frac{J_{2 i K }(2 \lambda)}{J_{-2 i K }(2 \lambda)}   \right|^2.  
\label{refCoeffSE} 
\end{align} 
Note that if $\lambda$ is real, then $R \equiv 1$ as it should with no absorption.

\section{\label{doubleExpAppend}Reflection Coefficient for Double-Exponential CAP} 

As in App.~\ref{expAppend}, we start from a unitless Schr\"odinger equation,
\begin{equation} 
   \left[ - \frac{d^2}{d z^2} -  \lambda_1^2 e^{ -z/2} 
   -i  \lambda_2^2 e^{-z} \right] \psi 
   =  K^2 \psi. 
\end{equation} 
with
\begin{equation} 
  z= \frac{x}{\beta}\qquad \lambda_1 = \alpha_1 \beta \qquad \lambda_2 = \alpha_2 \beta \qquad
K = k \beta.
\end{equation}
We assume both $\lambda_i$ are real, making the longer-range potential real in accord with
the discussion in the text.  That is, the real potential accelerates the wave before it encounters
the absorbing potential.

Defining
\begin{equation}
  \xi= 2 \lambda_2 e^{-\frac{z}{2}} \qquad \text{and}\qquad
  \varLambda = \frac{\lambda_1^2}{\lambda_2},  
\end{equation} 
we get 
\begin{equation} 
 \left[ \xi^2 \frac{d^2}{d \xi^2}+ \xi \frac{d}{d \xi} + 2\varLambda \xi + i \xi^2 + 4 K^2  \right] \psi =0 
\end{equation} 
This form of the equation makes clear the motivation for our choice of potential: having one
potential being the square of the other (in form) produces the polynomial in $\xi$ seen in the
equation. Since the polynomial is quadratic, the equation has analytic solutions.

Setting
$\gamma = e^{{i\pi}/{4}}$ and $\eta = \frac{1}{2}-\gamma \varLambda$,  the
solution can be written as
\begin{multline} 
 \psi = e^{\gamma^3 \xi + 2 i K \ln\xi} 
 \biggl[ C \, U(\eta+2iK, 1+4iK, -2\gamma^3 \xi) \\ +D \, L_{-\eta-2iK}^{4 i K}(-2 \gamma^3 \xi ) \biggr], 
\label{de_wfn}
\end{multline} 
where $U$ and $L$ are the confluent hypergeometric and Laguerre functions, respectively. 

Analyzing the asymptotic behavior, we have 
\begin{align} 
 e^{\gamma^3 \xi + 2 i K \ln \xi}  U \xrightarrow[z \rightarrow \infty]{}& \frac{(2\lambda_2)^{2iK} \Gamma(-4iK)}{\Gamma(\eta - 2i K)} e^{-iKz} 
 \nonumber \\ 
 &+ \frac{(2 \lambda_2)^{-2iK}2^{-4iK} e^{-\pi K} \Gamma(4iK)}{\Gamma(\eta+2iK)}  e^{iKz} \nonumber \\ 
 e^{\gamma^3 \xi + 2 i K \ln \xi} L \xrightarrow[z \rightarrow \infty]{}& (2 \lambda_2)^{2 i K} L_{-\eta-2iK}^{4 i K} (0) e^{-iK z}.
\end{align} 

The boundary condition $\psi(x=0)=0$ gives 
\begin{align} 
 D = -\frac{U(\eta+2iK,1+4iK,-4 \lambda_2 \gamma^3)}{L_{-\eta-2iK}^{4iK}(-4\lambda_2 \gamma^3)}C. 
\end{align} 

The reflection coefficient can now be extracted, and a little algebra applied, to give 
\begin{align} 
  R = \left| \frac{_1F_1(\eta+2iK,1+4iK,-4\gamma^3 \lambda_2 )}{_1F_1(\eta-2iK, 1-4iK, -4 \gamma^3 \lambda_2)} \right|^2. 
\end{align} 
For a purely real $V$~({\em i.e.}, $\lambda_2 \rightarrow 0$), we recover $R=1$ as we should. 
For $\lambda_1=0$, $R$ reduces to the single-exponential expression 
Eq.~(\ref{refCoeffSE}) with $\lambda$ purely imaginary.

\section{Complex boundary condition} 
\label{BCAppend}

\subsection{Zero potential}

It is easiest to understand the effect of the complex boundary condition ($b$ is complex)
\begin{equation}
 \frac{1}{\psi} \frac{d\psi}{dx} = b
\end{equation}
from the free-particle equation
\begin{equation}
   - \frac{d^2}{d z^2} \psi =  K^2 \psi
\end{equation}
in the same unitless notation as in the previous appendices.  In this notation, we must require
\begin{equation}
 \frac{1}{\psi} \frac{d\psi}{dz} = B,
\label{UnitlessBC}
\end{equation}
where $B=b\beta$.  The solution is, as usual,
\begin{equation}
\psi=C e^{iKz}+De^{-iKz}.
\end{equation}
When combined with the boundary condition, one finds
\begin{equation}
R=\biggl|\frac{B+iK}{B-iK}\biggr|^2.
\end{equation}
As discussed in the text, this $R$ goes to zero at $K=K_0$ when $B=-iK_0$.  Physically, this 
condition corresponds to setting an outgoing-wave boundary condition (on the left boundary)
for an incident momentum $K_0$.  The boundary is thus perfectly transparent at this momentum,
but imperfectly so at other momenta.

\subsection{Effect on bound states}

It is natural to ask what effect such a boundary condition might have on the energies of
any bound states in the system. One general way to answer this question is to consider
an arbitrary potential at $z=0$ and write
\begin{equation}
\psi = 
 \begin{cases}
   A F(z) & z\leqslant z_0 \\
   C e^{-\kappa (z-z_0)}+D e^{\kappa (z-z_0)} & z > z_0
 \end{cases}
\label{BoundWfnBC}
\end{equation}
with $\kappa=\beta\sqrt{2 m |E|/\hbar^2}$ and
$F(z)$ the energy-dependent solution appropriate to the arbitrary potential 
satisfying the required boundary condition at $z=0$.  Although this specific description 
assumes a short-range potential, a similar argument can be made for the Coulomb potential.

The wave function in Eq.~(\ref{BoundWfnBC}) must satisfy the log-derivative boundary
condition from Eq.~(\ref{UnitlessBC}) at $z=z_1$\,---\,this is why the exponentially-growing solution
must be retained.  Imposing this boundary condition leads to the following transcendental
equation for the energy of the bound state:
\begin{equation}
F'(z_0)+\kappa F(z_0)=\frac{B+\kappa}{B-\kappa}
\bigl[F'(z_0)-\kappa F(z_0)\bigr]
e^{-2\kappa(z_1-z_0)}.
\label{BoundEnergyEqnBC}
\end{equation}
This equation should be compared with the physical quantization condition
\begin{equation}
F'(z_0)+\kappa F(z_0)=0,
\label{BoundEnergyEqnPhys}
\end{equation}
to which Eq.~(\ref{BoundEnergyEqnBC}) reduces in the $z_1\rightarrow\infty$ limit as it should.

Since in any practical numerical solution of the TDSE the boundary of the numerical grid must 
be large compared to the size of the bound state, we will have $z_1\gg z_0$.  Therefore,
so long as $B-\kappa\neq 0$, the exponential term on the right-hand-side of Eq.~(\ref{BoundEnergyEqnBC})
will dominate\,---\,and thus make Eq.~(\ref{BoundEnergyEqnBC}) a very good approximation to 
Eq.~(\ref{BoundEnergyEqnPhys})\,---\,guaranteeing that the real part of the energy found with
the complex boundary condition will be very close to the physical energy.  To be sure, it will
acquire a small imaginary part reflecting the decay of the ground state due to the complex
boundary condition, but it should be completely negligible.

This result for the bound-state energies is completely consistent with the intuitive notion
that the effect of the complex boundary condition\,---\,indeed the CAPs, too\,---\,on the
bound states should be negligible so long as the bound-state wave function is vanishingly
small at the boundary $z=z_1$ (or in the absorbing region of the CAP).

\subsection{Single- and double-exponential CAPs}

The reflection coefficient for the single-exponential CAP with a complex boundary condition 
is still analytical.  It is found by imposing Eq.~(\ref{UnitlessBC}) on the general solution for the
single-exponential CAP from Eq.~(\ref{se_wfn}).  After a little algebra, one obtains
\begin{equation}
R = e^{4 K \arg \lambda^2} 
\biggl| \frac{(B+iK) J_{2iK}(2\lambda)-\lambda J_{1+2iK}(2\lambda)}
             {(B-iK) J_{-2iK}(2\lambda)-\lambda J_{1-2iK}(2\lambda)} \biggr|^2.
\label{seR_BC}
\end{equation}


The same procedure can be carried out for the double-exponential CAP using the general
solution from Eq.~(\ref{de_wfn}) to find
\begin{widetext}
\begin{equation} 
  R = \left| \frac{(B \!+\!iK\!+\!\gamma^3 \lambda_2)\,_1F_1(\eta\!+\!2iK,1\!+\!4iK,-4\gamma^3 \lambda_2 )
                      -2 \gamma^3  \lambda_2 (\eta\!+\!2iK)\,_1F_1(\eta\!+\!2iK\!+\!1,2\!+\!4iK,-4\gamma^3 \lambda_2 )/(1\!+\!4iK)}
                  {(B +iK+\gamma^3 \lambda_2)\,_1F_1(\eta-2iK, 1-4iK, -4 \gamma^3 \lambda_2)
                      -2iK\,_1F_1(\eta-2iK,-4iK,-4\gamma^3 \lambda_2 )} \right|^2. 
\label{deR_BC}
\end{equation} 
\end{widetext}


Note that both Eq.~(\ref{seR_BC}) and Eq.~(\ref{deR_BC}) reduce to the reflection coefficient
with $\psi=0$ in the $B\rightarrow\infty$ limit as they should.

\section{Reflection Coefficient for $-i \alpha_2^2/x^2$ CAP} 
\label{DipoleCAP}

We start from the Schr\"odinger equation
\begin{equation} 
   \left[ - \frac{\hbar^2}{2m} \frac{d^2}{d x^2} - \frac{\hbar^2}{2m} \frac{ia^2+\frac{1}{4}}{x^2}
    \right] \psi 
   =  \frac{\hbar^2}{2m} k^2 \psi. 
\end{equation} 
Defining $z=kx$, we must solve
\begin{equation}
   \left[ - \frac{d^2}{d z^2} - \frac{ia^2+\frac{1}{4}}{z^2}
    \right] \psi 
   =   \psi. 
\end{equation}
The general solution is
\begin{equation}
\psi = \sqrt{z} \biggl[ C J_{a/\gamma}(z)+ D J_{-a/\gamma}(z) \biggr]
\end{equation}
with $\gamma$=$e^{i\pi/4}$ as before.  Since we require $\psi(0)$=0, we need the
small-$z$ behavior of the Bessel functions
\begin{equation}
J_\nu(z) \xrightarrow[z\rightarrow 0]{} \frac{1}{\Gamma(1+\nu)}\biggl( \frac{z}{2} \biggr)^\nu.
\label{BesselSmallArg}
\end{equation}
For the general case of complex $a$, $a=a_r+i a_i$, one can show that requiring
the real part of the potential to be attractive and the imaginary part to be absorbing
leads to 
\begin{equation}
a_r > 0 \text{ and } a_i <0 \text{ with } |a_r|>|a_i|.
\end{equation}
These conditions, together with Eq.~(\ref{BesselSmallArg}), require us to set $D$=0.

Finally, using the large-$z$ behavior of the Bessel function,
\begin{equation}
\sqrt{z} J_{a/\gamma}(z) \xrightarrow[z\rightarrow \infty]{} \sqrt{\frac{2}{\pi}} 
                                     \cos\biggl(\frac{\pi}{4}+\frac{a\pi}{2\gamma}-z\biggr),
\end{equation}
allows us to extract the reflection coefficient,
\begin{equation}
R=e^{-\sqrt{2}\pi (a_r-a_i)}.
\end{equation}
Keeping in mind that $a_i<0$ under the conditions we have assumed, this equation shows
that both the real and imaginary parts of the CAP contribute to decreasing the final absorption.
We also see that this equation is identical to Eq.~(\ref{ds_HighEnergyR}) once the conversion
from $a$ to $\lambda_2$ is made.

\subsection*{Effect of truncation}

If we are willing to sacrifice the energy independence of $R$ at small energies by truncating the CAP at $x_0$,
\begin{equation}
V = 
 \begin{cases}
   - \frac{\hbar^2}{2m} \frac{ia^2+\frac{1}{4}}{x^2} & x\leqslant x_0 \\
   0 & x > x_0
 \end{cases},
\end{equation}
then one can again obtain an analytic expression for the reflection coefficient.

The wave function is
\begin{equation}
\psi =
 \begin{cases}
   A F(z) & z \leqslant z_0 \\
   C e^{-i (z-z_0)}+D e^{i (z-z_0)} & z > z_0
 \end{cases}
\end{equation}
with $z_0=kx_0$ and $F(z)=\sqrt{z} J_{a/\gamma}(z)$.  The reflection coefficient is thus
\begin{equation}
R = \biggl| \frac{D}{C} \biggr|^2 = \biggl| \frac{F(z_0)-i F'(z_0)}{F(z_0)+i F'(z_0)} \biggr|^2 .
\label{TruncDipoleR}
\end{equation}
The notation $F'$ indicates a derivative with respect to $z$, $F'=dF/dz$, and
the log-derivative of $F$ at $z_0$ is
\begin{equation}
\frac{F'}{F}\biggr|_{z=z_0} = \frac{1-2 a/\gamma}{2 z_0}+\frac{J_{a/\gamma-1}(z_0)}{J_{a/\gamma}(z_0)}.
\end{equation}

A plot of the reflection coefficient from Eq.~(\ref{TruncDipoleR}) looks qualitatively like the
double-sinh reflection coefficients in Figs.~\ref{ds} and \ref{RC}.  However, instead
of falling exponentially with $k$ at small $k$, it falls more slowly\,---\,like $1/z_0^4=1/(kx_0)^4$.
In addition, because of the sharp cutoff in the potential, $R$ oscillates in $z_0$ with minima
separated by $\pi$ at positions corresponding roughly to $z_0=kx_0=n\pi$.

\end{appendices}

\begin{acknowledgments} 
This work is supported by the Chemical Sciences, Geoscience, and Biosciences Division, Office for 
Basic Energy Sciences,Office of Science, U.S. Department of Energy. 
\end{acknowledgments} 
 
\bibliography{CAP}

\end{document}